\documentstyle[12pt]{article}
\begin{document}

\renewcommand{\theequation}{\thesection.\arabic{equation}}
\newcounter{saveeqn}
\newcommand{\add}{\addtocounter{equation}{1}}
\newcommand{\alpheqn}{\setcounter{saveeqn}{\value{equation}}%
\setcounter{equation}{0}%
\renewcommand{\theequation}{\mbox{\thesection.\arabic{saveeqn}{\alph{equation}}}}}
\newcommand{\reseteqn}{\setcounter{equation}{\value{saveeqn}}%
\renewcommand{\theequation}{\thesection.\arabic{equation}}}
\newenvironment{nedalph}{\add\alpheqn\begin{eqnarray}}{\end{eqnarray}\reseteqn}
\newsavebox{\PSLASH}
\sbox{\PSLASH}{$p$\hspace{-1.8mm}/}
\newcommand{\PS}{\usebox{\PSLASH}}
\newsavebox{\notrightarrow}
\sbox{\notrightarrow}{$\to$\hspace{-4mm}/}
\newcommand{\notto}{\usebox{\notrightarrow}}
\newsavebox{\PARTIALSLASH}
\sbox{\PARTIALSLASH}{$\partial$\hspace{-2.3mm}/}
\newcommand{\PARTIALS}{\usebox{\PARTIALSLASH}}
\newsavebox{\ASLASH}
\sbox{\ASLASH}{$A$\hspace{-2.1mm}/}
\newcommand{\AS}{\usebox{\ASLASH}}
\newsavebox{\KSLASH}
\sbox{\KSLASH}{$k$\hspace{-1.8mm}/}
\newcommand{\KS}{\usebox{\KSLASH}}
\newsavebox{\LSLASH}
\sbox{\LSLASH}{$\ell$\hspace{-1.8mm}/}
\newcommand{\LS}{\usebox{\LSLASH}}
\newsavebox{\QSLASH}
\sbox{\QSLASH}{$q$\hspace{-1.8mm}/}
\newcommand{\QS}{\usebox{\QSLASH}}
\newsavebox{\DSLASH}
\sbox{\DSLASH}{$D$\hspace{-2.8mm}/}
\newcommand{\DS}{\usebox{\DSLASH}}
\newsavebox{\DbfSLASH}
\sbox{\DbfSLASH}{${\mathbf D}$\hspace{-2.8mm}/}
\newcommand{\DBFS}{\usebox{\DbfSLASH}}
\newsavebox{\DELVECRIGHT}
\sbox{\DELVECRIGHT}{$\stackrel{\rightarrow}{\partial}$}
\newcommand{\PARVECR}{\usebox{\DELVECRIGHT}}
\thispagestyle{empty}
\begin{flushright}
IPM/P-2003/034
\end{flushright}
\vspace{0.5cm}
\begin{center}
{\large\bf{Planar and Nonplanar Konishi Anomalies and}\\
{\large\bf Effective Superpotential for Noncommutative
${\cal{N}}=1$
Supersymmetric $U(1)$}}\\
\vspace{1cm} {\bf Farhad Ardalan} \hspace{0.2cm} and
\hspace{0.2cm}{\bf
N\'eda Sadooghi}  \\
\vspace{0.5cm}
{\sl Department of Physics, Sharif University of Technology}\\
{\sl P.O. Box 11365-9161, Tehran-Iran}\\
and\\
{\sl Institute for Studies in Theoretical Physics and Mathematics (IPM)}\\
{\sl{School of Physics, P.O. Box 19395-5531, Tehran-Iran}}\\
{\it E-mails: ardalan, sadooghi@theory.ipm.ac.ir}
\end{center}
\vspace{0cm}
\begin{center}
{\bf {Abstract}}
\end{center}
\begin{quote}
The Konishi anomalies for noncommutative ${\cal{N}}=1$
supersymmetric $U(1)$ gauge theory arising from planar and
nonplanar diagrams are calculated. Whereas planar Konishi anomaly
is the expected $\star$-deformation of the commutative anomaly,
nonplanar anomaly reflects the important features of nonplanar
diagrams of noncommutative gauge theories, such as UV/IR mixing
and the appearance of nonlocal open Wilson lines. We use the
planar and nonplanar Konishi anomalies to calculate the effective
superpotential of the theory. In the limit of vanishing $|\Theta
p|$, with $\Theta$ the noncommutativity parameter, the
noncommutative effective superpotential depends on a gauge
invariant superfield, which includes supersymmetric Wilson lines,
and has nontrivial dependence on the gauge field supermultiplet.
\end{quote}
\newpage
\newpage
\subsubsection*{1\hspace{0.3cm}Introduction}
\setcounter{section}{1} \setcounter{equation}{0}

Quantum Field Theories with ${\cal{N}}=1$ supersymmetry have been
the subject of intense studies in the past 20 years. In particular
it has been shown that in many cases the exact form of the
effective superpotential can be determined using kinematical
constraints such as holomorphy and various symmetries, and also
approximate dynamical information about the asymptotic behavior of
the superpotential \cite{vy}-\cite{peskin}.

Recently a new technique for evaluating the effective
superpotential has been developed by Dijkgraaf and Vafa \cite{dv}.
They conjectured that the exact effective superpotential for
${\cal{N}}=1$ supersymmetric gauge theories can be constructed
from an associated matrix model, and, that the diagrams relevant
to the  effective theory are indeed planar.  Motivated by
Dijkgraaf-Vafa's conjecture, the authors of \cite{cachazo} derived
the effective superpotential using a certain generalized Konishi
anomaly.  After this work many other authors have analyzed
supersymmetric gauge theories along the line of Konishi anomaly
method.

In a separate development \cite{ooguri}, Ooguri and Vafa
considered a novel deformation of ${\cal{N}}=1$ supersymmetric
gauge theories in four dimensions, which involves a non-vanishing
anticommutation relation of fermionic coordinates of the
superspace $\{\theta^{\alpha}, \theta^{\beta}\}=C^{\alpha\beta}$,
(see also \cite{onehalf}). The consequences of this deformation
have been explored further by Seiberg \cite{seiberg2-2}. In
particular, it is shown,  that in the resulting ${\cal{N}}=1/2$
theory the ordinary space-time coordinates $x$ do not commute,
when the supercoordinate $\theta$ satisfy the anticommutation
relation $\{\theta^{\alpha}, \theta^{\beta}\}=C^{\alpha\beta}$. In
\cite{ooguri} Ooguri and Vafa have shown that the exact
superpotential of this so called $C$--deformed gauge theory
involves nonplanar diagrams. This theory has been intensely
studied recently  \cite{rey1, rey2}. In \cite{rey2}, the one-loop
effective superpotential of an ${\cal{N}}=1/2$ holomorphic
Wess-Zumino model is calculated, where both fermionic and bosonic
coordinates are made non(anti)commutative, and it is shown that
planar and nonplanar contributions exhibit different behavior.
Whereas planar diagrams yield an effective superpotential
proportional to $\Phi\star\log\Phi$, nonplanar diagrams are UV
divergent when bosonic noncommutativity is turned off. Here $\Phi$
is the Wess-Zumino hypermultiplet. Resumming the nonplanar
diagrams, they are expressed as a $\star$-product including open
Wilson lines in superspace. These supersymmetric open Wilson lines
are indeed responsible for noncommutative UV/IR mixing
\cite{minwalla}.

In this paper, we consider noncommutativity of space coordinates
only  \cite{nekrasov} and determine the exact effective
superpotential for the ${\cal{N}}=1$ supersymmetric $U(1)$ gauge
theory with particular emphasis on the role of nonplanar diagrams.
To find the exact
 effective superpotential, we follow the method of Konishi anomaly
\cite{vyt, seiberg} without recourse to the matrix model. A matrix
model formulation of noncommutative supersymmetric theories with
bosonic noncommutativity has been presented  in \cite{kawai},
where a $\star$-deformed Konishi anomaly is used to study
Dijkgraaf and Vafa's conjecture. We will show, using the previous
results on anomaly in noncommutative gauge theories \cite{ned1,
ned2, anomaly}, that such a $\star$-deformed Konishi anomaly
arises only from planar diagrams \cite{grisaru}. The theory,
however, has both planar and nonplanar diagrams. Nonplanar
diagrams do not contribute to the anomaly for large
noncommutativity parameter $\Theta$. For small $\Theta$, though,
they yield a finite contribution to the anomaly and have to be
taken into account. Indeed, nonplanar anomaly seems to be the
playing ground for many important features of noncommutative gauge
theories discovered in the last few years.

This paper is organized as follows: As most of the intricacies of
the effective action is in the form of the anomaly, we briefly
review the results of \cite{ned2} on planar and nonplanar
anomalies of noncommutative field theories in Section 2. After
deriving the global Noether current of the theory, we compute,
using Fujikawa's path integral method \cite{fujikawa}, the gauge
covariant (planar) and invariant (nonplanar) anomalies
corresponding to the covariant and invariant currents of $U(1)$
gauge theory. We will show that due to the UV/IR mixing
\cite{minwalla}, the nonplanar anomaly arises as a singularity at
the limit of zero $|\Theta p|$ where $\Theta$ is the
noncommutativity parameter and $p$ the momentum. In Section 3, we
will then calculate the covariant (planar) and invariant
(nonplanar) Konishi anomalies using a supersymmetric version of
Fujikawa's path integral method, originally introduced by Konishi
\cite{konishi}. To obtain $\star$-gauge invariant expressions for
the invariant (nonplanar) Konsihi anomaly, it is attached to a
noncommutative supersymmetric open Wilson line, defined by the
noncommutative generalization of supersymmetric Wilson lines
\cite{mansouri}.

There are three ingredients which are necessary for evaluating the
exact effective superpotential. The first is the Konsihi anomaly,
whose nontrivial structure will be presented in Section 3. The
second is the one-loop $\beta$-function of the theory. In section
4, the background field method will be used to determine the
one-loop $\beta$-function of noncommutative ${\cal{N}}=1$
supersymmetric $U(1)$ gauge theory with $N_{f}$ flavor pairs in
the fundamental and antifundamental representation. This method
has been previously used in \cite{khoze} and \cite{alvarez}, where
the $\beta$-function of nonsupersymmetric field theory was
determined and was shown that it exhibits a UV/IR mixing. In the
supersymmetric generalization, we will show that in the
$|p|\gg\frac{1}{\sqrt{\Theta}}$ limit only planar diagrams
contribute to the one-loop $\beta$-function, and that the theory
is asymptotically free for $N_{f}< 3$. In the limit of
$|p|\ll\frac{1}{\sqrt{\Theta}}$, however, where the nonplanar
diagrams also contribute, the theory is IR free for any number of
flavors.

Finally, the third ingredient is the anomaly corresponding to the
$U_{R}(1)$ symmetry of the theory. As in the ordinary
supersymmetric QCD, it receives contributions from the fermionic
fields of the $N_{f}$ matter supermultiplets and also from the
gaugino field in the gauge supermultiplet in the adjoint
representation.  We will calculate the anomaly corresponding to
the R-symmetry separately using the Fujikawa Method and will show,
that, due to UV/IR mixing, the anomaly arising from $N_{f}$
flavors in the matter supermultiplet in the fundamental
representation receives contribution only in the limit of
vanishing  $|\Theta p|$, whereas the contribution of the gaugino
in the adjoint representation to this R-anomaly appears for any
$|\Theta p|$.
\par
Having these special ingredients of noncommutative supersymmetric
gauge theories on hand, in Section 4 we follow the conventional
methods of ordinary SQCD, outlined in Refs. \cite{vy,
vyt,seiberg2} (for a recent review see \cite{terning}), to find
the effective superpotential of the gaugino condensate for two
different limits of vanishing and arbitrary but finite $|\Theta
p|$, separately.  We also determine the Affleck-Dine-Seiberg meson
field superpotential \cite{seiberg1} for the noncommutative theory
in these two limits. Section 5 is devoted to discussions.
\subsubsection*{2\hspace{0.3cm}Anomalies in Noncommutative $U(1)$ Gauge Theory}
\setcounter{section}{2} \setcounter{equation}{0} To begin we will
briefly review the results of Refs. \cite{ned1, ned2}. The
important observations there were that a noncommutative gauge
theory with
 matter fields in fundamental representation consists of two different
  vector currents and axial vector currents and that the anomalies corresponding
  to these axial vector currents arise from planar and nonplanar diagrams of
  the noncommutative gauge theory. Whereas the planar anomaly \cite{ned1},
  arising from planar diagrams, is the $\star$-deformation of the commutative
  Adler-Bell-Jackiw (ABJ) anomaly \cite{adler}, the nonplanar anomaly reflects
  the unconventional and important behavior of nonplanar diagrams of
  noncommutative gauge theories. There are at least two important
  features related to the nonplanar anomaly of noncommutative gauge theories:

The first property is the UV/IR mixing indicating a certain
singularity at $|\Theta p|\to 0$ limit. The second important point
related to the nonplanar anomaly is the appearance of a
generalized $\star$-product between $F_{\mu\nu}$ and its dual. In
\cite{ned2}, we calculated perturbatively the contribution of
nonplanar diagrams to the axial anomaly using two different
regularization methods; dimensional and Pauli-Villars
regularizations, and showed that the result for the nonplanar
anomaly is not $\star$-gauge invariant if we consider only the
triangle, square and pentagon diagrams as in the ordinary
commutative non-Abelian gauge theories. To restore the
$\star$-gauge invariance, it is necessary to attach $F_{\mu\nu}$
and its dual to an open Wilson line with a length proportional to
$\sqrt{\Theta}$. The expansion of the Wilson line in the external
gauge fields produce infinitely many diagrams whose contributions
guarantee the gauge invariance of the final result
\cite{starprime,startrek}. The invariant anomaly in noncommutative
gauge theories has been also studied in a number of other works
\cite{intriligator,armoni,banerjee}, and there are certain
disagreements in the literature; here we will reproduce our
previous results \cite{ned2} by the Fujikawa method. But, when
using the result of the anomaly for determination of the effective
action of the gauginos in the supersymmetric theory we will not
commit ourselves to a particular form of the anomaly, $S'$ in this
paper. Any of the forms found for $S'$ in the literature may be
substituted in the action.\footnote{The vanishing anomaly of
\cite{intriligator} is in fact excluded by the arguments of
\cite{armoni}: The non-vanishing of the covariant anomaly forces a
nonzero value for the invariant anomaly (see the discussion in
section 2.3).}
\par
In this section we first derive the global Noether currents of
noncommutative nonsupersymmetric $U(1)$ gauge theory and resolve
the ambiguity which arises in determining these currents. We then
calculate the covariant (planar) and invariant (nonplanar)
anomalies in noncommutative $U(1)$ gauge theory using Fujikawa's
path integral method \cite{fujikawa}, which can be easily
generalized to the Konishi anomaly of supersymmetric gauge
theories (see section 3).
\par
Let us begin by fixing our notations and by recalling that
noncommutative gauge theory is characterized by replacing the
familiar product of functions with the $\star$-product
\begin{eqnarray}\label{NS1}
f(x)\star g(x)\equiv f(x+\xi)\ \exp\left(\frac{i\Theta^{\mu\nu}}{2}\
\frac{\partial}{\partial\xi^{\mu}}\frac{\partial}{\partial\zeta^{\nu}}\right) g(x+\zeta)\Bigg|_{\xi=\zeta=0},
\end{eqnarray}
where $\Theta^{\mu\nu}$ is a real antisymmetric matrix, and reflects the
noncommutativity of the coordinates
\begin{eqnarray}\label{NS2}
\big[x^{\mu},x^{\nu}\big]=i\Theta^{\mu\nu}.
\end{eqnarray}
The action of the noncommutative $U(1)$ gauge theory with matter fields
in the fundamental representation is
\begin{eqnarray}\label{NS3}
{\cal{S}}[A_{\mu},\bar{\psi},\psi]=-\frac{1}{4}\int d^{4}x\ F_{\mu\nu}\star F^{\mu\nu}+\int d^{4}x
\ \bar{\psi}(x)\star\left(i\DS-m\right)\psi(x),
\end{eqnarray}
with the field strength tensor
\begin{eqnarray}\label{NS4}
F_{\mu\nu}\equiv \partial_{\mu}A_{\nu}-\partial_{\nu}A_{\mu}+ig\big[A_{\mu},A_{\nu}\big]_{\star},
\end{eqnarray}
and the covariant derivative
\begin{eqnarray}\label{NS5}
D_{\mu}\psi(x)\equiv \partial_{\mu}\psi(x)+igA_{\mu}(x)\star \psi(x).
\end{eqnarray}
The above action is invariant under global transformation of the matter fields
\begin{eqnarray}\label{global}
\delta \psi(x)=i\alpha\psi(x),\hspace{1cm}\mbox{and}\hspace{1cm}\delta\bar{\psi}(x)=-i\alpha\bar{\psi}(x).
\end{eqnarray}
The corresponding Noether current is given unambiguously by
varying the Lagrangian density giving
\begin{eqnarray}
j^{\mu}_{inv.}(x)\equiv
 =\bar{\psi}(x)\gamma^{\mu}\star\psi(x),\hspace{1cm}\mbox{with}\hspace{1cm}\partial_{\mu}j^{\mu}_{inv.}(x)=0.
\end{eqnarray}
We should emphasize that the covariant current,
$J^{\mu}_{cov.}\equiv
\psi_{\beta}\star\bar{\psi}_{\alpha}(\gamma^{\mu})^{\alpha\beta}$,
is not conserved in the Noether sense, {\it i.e.}
$\partial_{\mu}J^{\mu}_{cov.}\neq 0$. It satisfies the equation
\begin{eqnarray}
D_{\mu}J^{\mu}_{cov.}=0.
\end{eqnarray}
Similar variational procedure leads to the anomalous global axial vector current
\begin{eqnarray}\label{NS9}
j^{\mu,5}_{inv.}(x)=-\bar{\psi}_{\alpha}(x)\star \psi_{\beta}(x)\ \left(\gamma_{\mu}\gamma_{5}\right)^{\alpha\beta},
\end{eqnarray}
resulting from the invariance of the action (\ref{NS3}) under the global
axial transformation $\delta\psi=i\alpha\gamma_{5}\psi$. A second axial vector current can also be defined
\begin{eqnarray}\label{NS10-c}
J^{\mu,5}_{cov.}(x)=\psi_{\beta}(x)\star\bar{\psi}_{\alpha}(x)\ \left(\gamma_{\mu}\gamma_{5}\right)^{\alpha\beta}.
\end{eqnarray}
These two currents satisfy $\partial_{\mu}j^{\mu,5}(x)=0$ and
$D_{\mu}J^{\mu,5}_{cov.}(x)=0$ in the chiral
limit.\footnote{Relations between different currents and their
connection to the commutative currents are discussed in
\cite{banerjee2, banerjee3}.}
\par
In \cite{ned1, ned2} we have shown, using diagrammatic methods, that the anomaly
 corresponding to the covariant axial vector current, $J_{cov.}^{\mu,5}(x)$, arises
 only from planar diagrams, whereas the anomaly corresponding to the invariant
 current, $j_{inv.}^{\mu,5}(x)$,  receives contributions only from the nonplanar
  diagrams. In the following  section, we derive anew
  our previous results using Fujikawa's path integral method \cite{fujikawa}.
\subsubsection*{2.1\hspace{0.3cm}Covariant (Planar) $U_{A}(1)$ Anomaly}\label{Sec21}
Consider the partition function of noncommutative $U(1)$ gauge theory
with matter fields in the fundamental representation
\begin{eqnarray}\label{NS10}
{\cal{Z}}=\int {\cal{D}}\psi\ {\cal{D}}\bar{\psi}\ e^{-iS_{F}[\psi,\bar{\psi}]},
\end{eqnarray}
where
\begin{eqnarray}\label{NS10-b}
S_{F}=i\int d^{4}x \bigg[\bar{\psi}_{\alpha}(x)\star(\gamma_{\mu})^{\alpha\beta}\partial_{\mu} \psi(x)+
ig \bar{\psi}_{\alpha}(x)\star A_{\mu}(x)(\gamma^{\mu})^{\alpha\beta}\star\psi_{\beta}(x)\bigg],
\end{eqnarray}
is the fermionic part of the action (\ref{NS3}) in the massless limit.
Requiring that ${\cal{Z}}$ remains invariant under the following
{\it fundamental} axial change of variables,
\begin{eqnarray}\label{NS11}
\delta_{5}\psi(x)=i\alpha(x)\gamma_{5}\star\psi(x),
\end{eqnarray}
the {\it covariant} ({\it planar}) axial anomaly can be determined. Under these local
change of variables  the fermionic part of the action transforms as
\begin{eqnarray}\label{NS12}
S_{F}\longrightarrow S'_{F}=S_{F}-\int d^{4}x\ D_{\mu}J^{\mu,5}_{cov.}(x)\star \alpha(x),
\end{eqnarray}
with the covariant current $J_{cov.}^{\mu,5}$ given in (\ref{NS10-c}). The covariant derivative $D_{\mu}$ is defined by
\begin{eqnarray}\label{NS13}
D_{\mu}J^{\mu,5}_{cov.}(x)\equiv \partial_{\mu}J^{\mu,5}_{cov.}+ig\big[A_{\mu},J_{cov.}^{\mu,5}\big]_{\star}.
\end{eqnarray}
To calculate the Jacobian of the transformations (\ref{NS11}), we
expand $\psi$ and $\bar{\psi}$ as a linear  combination of the
eigenfunctions $\varphi_{n}$ and $\varphi^{\dagger}_{n}$ of the
Dirac operator
\begin{eqnarray}\label{NS14}
\psi(x)=\sum\limits_{n} a_{n}\ \varphi_{n}(x),\hspace{1cm}\mbox{and}\hspace{1cm}\bar{\psi}(x)=\sum\limits_{n} b_{n}\
\varphi_{n}^{\dagger}(x),
\end{eqnarray}
where
\begin{eqnarray}\label{NS15}
a_{m}=\int d^{4}x\ \varphi_{m}^{\dagger}(x)\star \psi(x),
\end{eqnarray}
transforms as
\begin{eqnarray}\label{NS16}
a'_{m}=\sum\limits_{n}\left(\delta_{mn}+{\cal{C}}_{mn}\right)a_{n},
\hspace{1cm}\mbox{with}\hspace{1cm}{\cal{C}}_{nm}\equiv i\int d^{4}x\ \alpha(x)\star\varphi_{n,\beta}(x)
\star\varphi_{m,\alpha}^{\dagger}(x)(\gamma_{5})^{\alpha\beta}.
\end{eqnarray}
The Jacobian ${\cal{J}}$ therefore is
\begin{eqnarray}\label{NS18}
{\cal{J}}=\exp\left(\sum\limits_{n}{\cal{C}}_{nn}\right)=\exp\left(i\int d^{4}x\ \alpha(x)\star\sum\limits_{n}
\varphi_{n,\beta}(x)\star\varphi_{n,\alpha}^{\dagger}(x)(\gamma_{5})^{\alpha\beta}\right),
\end{eqnarray}
and the measure ${\cal{D}}\psi {\cal{D}}\bar{\psi}$
 transforms as
\begin{eqnarray}\label{NS19}
{\cal{D}}\psi\ {\cal{D}}\bar{\psi}\longrightarrow {\cal{D}}\psi'\ {\cal{D}}\bar{\psi'}=
\exp\left(-2i\int d^{4}x\ \alpha(x)\star\sum\limits_{n}\varphi_{n,\beta}(x)\star\varphi_{n,\alpha}^{\dagger}(x)
(\gamma_{5})^{\alpha\beta}\right){\cal{D}}\psi\ {\cal{D}}\bar{\psi}.
\end{eqnarray}
Combining the relations (\ref{NS12}) and (\ref{NS19}) we find that,
the partition function (\ref{NS10}) remains invariant if and only if
\begin{eqnarray}\label{NS20}
D_{\mu}J^{\mu,5}_{cov.}(x)=2\
\left(\sum\limits_{n}\varphi_{n,\beta}(x)\star\varphi_{n,\alpha}^{\dagger}(x)(\gamma_{5})^{\alpha\beta}\right).
\end{eqnarray}
As the $\varphi_{n}$'s transform in the fundamental
representation, it can be easily checked that both sides of the
above equation are covariant under $\star$-gauge transformation.
Here, as in the commutative case, the r.h.s. of (\ref{NS20}) must
be regulated, using a gauge covariant Gaussian damping factor
$\exp(-\frac{\DS^{\ 2}}{M^{2}})$ with $-\DS^{\
2}=-D^{2}-\frac{g}{2}\sigma_{\mu\nu}F^{\mu\nu}$,
$\sigma_{\mu\nu}=\frac{i}{2}\big[\gamma_{\mu},\gamma_{\nu}\big]$
and $M$ the regulator mass
\begin{eqnarray}\label{NS21}
D_{\mu}J^{\mu,5}_{cov.}(x)=\lim\limits_{M\to\infty}2\ \left(\sum\limits_{n}
\bigg[e^{\frac{-\DS^{\ 2}}{M^{2}}}\star\varphi_{n}(x)\bigg]_{\beta}
\star\varphi_{n,\alpha}^{\dagger}(x)(\gamma_{5})^{\alpha\beta}\right).
\end{eqnarray}
Transforming to the Fourier space and after some standard manipulations, we arrive at
\begin{eqnarray}\label{NS23}
D_{\mu}J^{\mu,5}_{cov.}(x) =2\lim\limits_{M\to
\infty}\mbox{tr}\left(\gamma_{5}\frac{1}{2!}\left(-\frac{g}{2M^{2}}
\sigma_{\mu\nu}F^{\mu\nu}(x)\right)^{2}_{\star}\right)\int
\frac{d^{4}k}{(2\pi)^{4}}\ e^{+\frac{k^{2}}{M^{2}}}+
{\cal{O}}(\frac{1}{M^{2}}).
\end{eqnarray}
Taking the limit $M\to \infty$ and using
tr$\left(\gamma_{5}\sigma_{\mu\nu}\sigma_{\lambda\rho}\right)=
4i\varepsilon_{\mu\nu\lambda\rho}$, the only remaining finite
term, the planar $U_{A}(1)$ anomaly is given by
\begin{eqnarray}\label{NS24}
D_{\mu}J^{\mu,5}_{cov.}(x)=-\frac{g^{2}}{16\pi^{2}} F_{\mu\nu}(x)\star \tilde{F}^{\mu\nu}(x),
\end{eqnarray}
and $\tilde{F}_{\mu\nu}\equiv \varepsilon_{\mu\nu\lambda\rho} F^{\lambda\rho}$.
The result  is indeed the expected $\star$-generalization of the Adler-Bell-Jackiw
anomaly (ABJ) \cite{adler}, in agreement with previous calculation \cite{ned1, anomaly}.
\subsubsection*{2.2\hspace{0.3cm}Noncommutative Open Wilson Line and
Generalized $\star$-Product}
In the next section we will calculate the $U(1)$ anomaly corresponding
to the invariant current of the theory. We will see that in order to
receive a gauge invariant expression for the so called invariant or
nonplanar anomaly, we have to attach the anomaly to an open Wilson line
with the length proportional to  $\sqrt{\Theta}$, where $\Theta$ is the
noncommutativity parameter. In this section we will review some general
aspects of noncommutative open Wilson line in the nonsupersymmetric case.
\par
The noncommutative open Wilson lines \cite{starprime}
 are noncom\-mutative gen\-er\-al\-iza\-tion of the com\-mutative
 Schwinger's line integrals, and are defined by
\begin{eqnarray}\label{NS27}
W(x,\ell)=P_{\star}\exp\left(i\int\limits_{0}^{1}d\sigma\
\frac{d\xi^{\mu}(\sigma)}{d\sigma}A_{\mu}(x+\xi(\sigma))\right)_{\star},
\end{eqnarray}
where $0\leq\sigma\leq 1$, $x$ is the basis of the Wilson line
 and $\ell$ is its length.
It transforms under the local gauge transformation as
\begin{eqnarray}\label{NS28}
W(x,\ell)\to U(x)\star W(x,\ell)\star U^{\dagger}(x+\ell),
\end{eqnarray}
with $U(x)\equiv e^{i\alpha(x)}$ and $\alpha(x)$ an arbitrary
function. Generally for a $\star$-gauge covariant operator
${\cal{O}}_{cov.}(x)$ transforming as
\begin{eqnarray}\label{NS29}
{\cal{O}}_{cov.}(x)\to U(x)\star{\cal{O}}_{cov.}(x)\star U^{\dagger}(x),
\end{eqnarray}
one defines a modified Fourier transformation
\begin{eqnarray}\label{NS30}
\tilde{\cal{O}}_{inv.}(k)=\int d^{4}x\ {\cal{O}}_{cov.}(x)\star W(x,\ell)\star e^{ikx}\equiv \int d^{4}x\
P_{\star}\bigg[{\cal{O}}_{cov.}(x)W(x,\ell)\bigg]\star e^{ikx},
\end{eqnarray}
which is  manifestly $\star$-gauge invariant, if the length
of the Wilson line is given by $\ell_{\mu}=\Theta_{\mu\nu}k^{\nu}$.
 The gauge invariant version of ${\cal{O}}_{cov.}$ is given therefore
 by going back to the coordinate space
\begin{eqnarray}\label{NS32}
{\cal{O}}_{inv.}(x)=\int \frac{d^{4}k}{(2\pi)^{4}} e^{-ikx}\int d^{4}y\
P_{\star}\bigg[{\cal{O}}_{cov.}(y)W(y,\ell=\Theta^{\mu\nu}k_{\nu})\bigg]\star e^{iky}.
\end{eqnarray}
It is clear that the first term in the expansion of the Wilson line
in terms of the small gauge field is the same gauge covariant operator
 ${\cal{O}}_{cov.}$
\begin{eqnarray}\label{NS33}
{\cal{O}}_{inv.}(x)={\cal{O}}_{cov.}(x)+\mbox{higher order terms}.
\end{eqnarray}
The situation changes if the gauge covariant operator is a product
of $n$ other gauge covariant operators. Suppose the operator
${\cal{Q}}_{cov.}$ is a product of $n$ operators ${\cal{O}}_{i}$
\begin{eqnarray}\label{NS34}
{\cal{Q}}_{cov.}(x)={\cal{O}}_{1}(x)\star{\cal{O}}_{2}(x)\star\cdots\star{\cal{O}}_{n}(x),
\end{eqnarray}
with
\begin{eqnarray}\label{NS35}
{\cal{O}}_{i}(x)\to U(x)\star{\cal{O}}_{i}(x)\star U^{\dagger}(x).
\end{eqnarray}
In this case there are two possibilities to attach the operators
to the Wilson line in order to build a gauge invariant operator.
The first possibility is to  attach ${\cal{Q}}_{cov.}(x)$ itself
at the end of the Wilson line.  Its gauge invariant version is
again given by (\ref{NS32})
\begin{eqnarray}\label{NS36}
{\cal{Q}}_{inv.}(x)=\int \frac{d^{4}k}{(2\pi)^{4}}\ e^{-ikx}\int d^{4}y\ P_{\star}
\bigg[{\cal{Q}}_{cov.}(y)W(y,\ell=\Theta^{\mu\nu}k_{\nu})\bigg]\star e^{iky}.
\end{eqnarray}
Expanding the Wilson line in terms of the external gauge field leads to
\begin{eqnarray}\label{NS37}
{\cal{Q}}_{inv.}(x)={\cal{O}}_{1}(x)\star{\cal{O}}_{2}(x)\star\cdots\star{\cal{O}}_{n}(x)+\mbox{
higher order terms}.
\end{eqnarray}
The second possibility is to smear the gauge covariant operators
${\cal{Q}}_{cov.}(x)$  along the Wilson line by attaching
${\cal{O}}_{i}(x), i=1,\cdots,n$ at different insertion points and
eventually integrating over all these insertion points
\cite{startrek}
\begin{eqnarray}\label{NS38}
\tilde{\cal{Q}}_{inv.}(k)&=&\int d^{4}x\  \left(\prod\limits_{i=1}^{n}\int\limits_{0}^{1}d\tau_{i}\right)\
P_{\star}\bigg[W(x,C)\prod\limits_{i=1}^{n}{\cal{O}}_{i}\left(x+\xi(\tau_{i})\right)\bigg]\star e^{ikx}\nonumber\\
&\equiv&\int d^{4}x\ L_{\star}\bigg[W(x,C)\prod\limits_{i=1}^{n}{\cal{O}}_{i}(x)\bigg]\star e^{ikx}.
\end{eqnarray}
The new path ordering $L_{\star}$, defined in the above equation,
expresses the smearing of $n$ covariant operators ${\cal{O}}_{i}$
along the Wilson line using new parameters $\tau_{i},
i=1,\cdots,n$. The gauge invariant version of
${\cal{Q}}_{cov.}(x)$ in the $x$ space can therefore be given by
an inverse Fourier transformation
\begin{eqnarray}\label{NS39}
{\cal{Q}}_{inv.}(x)=\int\frac{d^{4}k}{(2\pi)^{4}}\ e^{-ikx}\int d^{4}y\ L_{\star}
\left(W(x,C)\prod\limits_{i=1}^{n}{\cal{O}}_{i}(x)\right)\star e^{iky}.
\end{eqnarray}
The Wilson line appearing in the above equation can be expanded
again in powers of the external gauge fields $A_{\mu}$. The crux
of this expansion is that here, in contrast to the previous case
(\ref{NS33}), the first term of the expansion will include a
generalized $\star$-product \cite{starprime}, $\star_{n}$, between
$n$ gauge covariant operators
\begin{eqnarray}\label{NS40}
{\cal{Q}}_{inv.}(x)=\big[{\cal{O}}_{1}(x),\cdots,{\cal{O}}_{n}(x)\big]_{\star_{n}}+\mbox{higher order terms}.
\end{eqnarray}
For $n=2$, (\ref{NS40}) reads
\begin{eqnarray}\label{NS41}
{\cal{Q}}_{inv.}(x)={\cal{O}}_{1}(x)\star' {\cal{O}}_{2}(x)+\mbox{higher order terms},
\end{eqnarray}
with $\star'$-product given by
\begin{eqnarray}\label{x2-1}
f(x)\star' g(x)\equiv
f(x+\xi)\frac{\sin\left(\frac{\Theta_{\mu\nu}}{2}\
\frac{\partial}{\partial\xi}\frac{\partial}{\partial\zeta}\right)}{\frac{\Theta_{\mu\nu}}{2}\
\frac{\partial}{\partial\xi}\frac{\partial}{\partial\zeta}}
g(x+\zeta)\Bigg|_{\xi=\zeta=0}.
\end{eqnarray}
We will now calculate the anomaly corresponding to gauge invariant
current, where the open Wilson lines are introduced to provide the
gauge invariance of the result.
\subsubsection*{2.3\hspace{0.3cm}Invariant (Nonplanar) $U_{A}(1)$ Anomaly}
We will now use the invariance of the partition function
(\ref{NS10}) under the  {\it antifundamental} local axial change
of variables
\begin{eqnarray}\label{NS44}
\delta_{5}\psi(x)=i\psi(x)\star\alpha(x)\gamma_{5},
\end{eqnarray}
to obtain the {\it invariant} ({\it nonplanar}) anomaly. Under
this change of variable the fermionic part of the action
(\ref{NS10-b}) transforms as
\begin{eqnarray}\label{NS45}
S_{F}\longrightarrow S'_{F}=S_{F}-\int d^{4}x\ \partial_{\mu}j^{\mu,5}_{inv.}(x)\star \alpha(x),
\end{eqnarray}
with
$J_{inv.}^{\mu,5}=-\bar{\psi}_{\alpha}\star\psi_{\beta}\left(\gamma_{\mu}\gamma_{5}\right)^{\alpha\beta}$
given in (\ref{NS9}), and we get
\begin{eqnarray}\label{NS46}
{\cal{D}}\psi\ {\cal{D}}\bar{\psi}\to {\cal{D}}\psi'\ {\cal{D}}\bar{\psi'}=\exp\left(-2i\int d^{4}x\
\alpha(x)\star\sum\limits_{n}\varphi_{n,\alpha}^{\dagger}(x)\star(\gamma_{5})^{\alpha\beta}\varphi_{n,\beta}(x)\right)\
{\cal{D}}\psi\ {\cal{D}}\bar{\psi},
\end{eqnarray}
for the Jacobian of the transformation. The condition for the invariance of the partition function now reads
\begin{eqnarray}\label{NS47}
\partial_{\mu}j_{inv.}^{\mu,5}(x)=2\ \left(\sum\limits_{n}\varphi_{n, \alpha}^{\dagger}(x)
\star(\gamma_{5})^{\alpha\beta}\varphi_{n,\beta}(x)\right).
\end{eqnarray}

Note that since the l.h.s. of this equation is gauge invariant a
gauge invariant version of Gaussian damping factor
$\exp(-\DS^{2}/M^{2})=\exp\left(-D^{2}/M^{2}-g\
\sigma_{\mu\nu}F^{\mu\nu}/2\right)$ has to be used to regularize
the r.h.s. To do this the field tensor $F^{\mu\nu}$ is to be
smeared along an open Wilson line $W(y,C)$ with the length
$C$.\footnote{The calculation of the unintegrated form of the
invariant anomaly using the Seiberg-Witten map \cite{banerjee}
suggests the smearing procedure used here.} We obtain
\begin{eqnarray}\label{x3}
\lefteqn{
\partial_{\mu}j_{inv.}^{\mu,5}(x)=\lim\limits_{M\to \infty}\
2\int\frac{d^{4}p}{(2\pi)^{4}}\ e^{-ipx}\int d^{4}y\
\sum\limits_{n}\bigg[e^{-\frac{D^{2}}{M^{2}}}\ \varphi^{\dagger}_{n,\alpha}(y)
}\nonumber\\
&& \times (\gamma_{5})^{\alpha\beta}\star P_{\star}
\left(W(y,C)\exp\left(-\frac{g}{2M^{2}}\int\limits_{0}^{1} d\tau\
F_{\mu\nu}(y+\tilde{p}\tau)\sigma^{\mu\nu}\right)\right)_{\beta}^{\ \delta}\star\varphi_{n,\delta}(y)
\bigg] \star e^{ipy}.
\end{eqnarray}
Expanding now the exponential including the field strength tensor and using
the completeness of the basis functions $\tilde{\varphi}_{n}(k)$
\begin{eqnarray*}
\sum\limits_{n}\tilde{\varphi}_{n,\alpha}^{\dagger}(k)\tilde{\varphi}_{n,\beta}(k')=
(2\pi)^{4}\delta(k-k')\delta_{\alpha\beta},
\end{eqnarray*}
we find that the only remaining term in the limit $M\to \infty$ is
given by
\begin{nedalph}\label{NS51a}
\partial_{\mu}j^{\mu,5}_{inv.}(x)={\cal{A}}_{inv.}^{nonplanar}(x),
\end{eqnarray}
where
\begin{eqnarray}\label{NS51b}
\lefteqn{ {\cal{A}}_{inv.}^{nonplanar}(x)=\lim\limits_{M\to
\infty}\ 2\
\mbox{tr}\left(\gamma^{5}\sigma^{\mu\nu}\sigma^{\rho\lambda}\right)\
\left(-\frac{g}{2M^{2}}\right)^{2}\frac{1}{2!}\
\int\frac{d^{4}p}{(2\pi)^{4}}\ e^{-ipx}\ \int
\frac{d^{4}k}{(2\pi)^{4}} e^{+\frac{k^{2}}{M^{2}}}
}\nonumber\\
&&\times \int d^{4}y\  e^{-iky}\ \star P_{\star}\left(W(y)\star\int\limits_{0}^{1}d\tau_{1}
\int\limits_{0}^{1}d\tau_{2}\ F_{\mu\nu}(y+\tilde{p}\tau_{1})\ F_{\rho\lambda}(y+\tilde{p}\tau_{2})\right)
\star e^{iky}\star e^{ipy}\nonumber\\
&=&\lim\limits_{M\to \infty}\frac{ig^{2}}{M^{4}}\int\frac{d^{4}p}{(2\pi)^{4}}\ e^{-ipx}\ \int
 \frac{d^{4}k}{(2\pi)^{4}} e^{\frac{k^{2}}{M^{2}}}\nonumber\\
&&\times\int d^{4}y\ P_{\star}\left(
W(y-\tilde{k})\star\int\limits_{0}^{1}d\tau_{1}\int\limits_{0}^{1}d\tau_{2}\
F_{\mu\nu}(y-\tilde{k}+\tilde{p}\tau_{1})\
F_{\rho\lambda}(y-\tilde{k}+\tilde{p}\tau_{2})\right)\star
e^{ipy}.\nonumber\\
\end{nedalph}
\\
Here  the relation
\begin{eqnarray}\label{n1}
f(y)\star e^{iky}=e^{iky}\star f(y-\tilde{k}),
\end{eqnarray}
with $\tilde{k}_{\mu}=\Theta_{\mu\nu}k^{\nu}$ is used.
\par
In the following we will calculate the first term in the expansion
of the open Wilson line in the powers of external gauge field and
we will show that as from the perturbative calculation performed
in \cite{ned2}, a generalized $\star'$-product will emerge between
the two field strength tensors. Taking (\ref{NS51b}) and expanding
the Wilson line in the orders of small external gauge field, we
arrive at
\begin{eqnarray}\label{NS52}
\lefteqn{{\cal{A}}_{inv.}^{nonplanar}(x)\Bigg|_{first\ term}=
\lim\limits_{M\to \infty}+\frac{ig^{2}}{M^{4}}\int
\frac{d^{4}k}{(2\pi)^{4}}\ e^{+\frac{k^{2}}{M^{2}}}\
\int \frac{d^{4}p}{\left(2\pi\right)^{4}}\ e^{-ipx}
}
\nonumber\\
&&\hspace{0cm}\times\int d^{4}y\
P_{\star}\left(\int\limits_{0}^{1}d\tau_{1}\int\limits_{0}^{1}d\tau_{2}\
F_{\mu\nu}(y-\tilde{k}+\tau_{1}\tilde{p})\star
\tilde{F}^{\mu\nu}(y-\tilde{k}+\tau_{2}\tilde{p})\right)\star
e^{ipy},
\end{eqnarray}
with
\begin{eqnarray*}
\lefteqn{\int\limits_{\tau_{0}}^{\tau}d\tau_{1}\
\int\limits_{\tau_{0}}^{\tau_{1}}d\tau_{2}\
P_{\star}\left({\cal{O}}_{1}(x+\xi(\tau_{1}))\
{\cal{O}}_{2}(x+\xi(\tau_{2}))\right)\equiv}\nonumber\\
&&\int\limits_{\tau_{0}}^{\tau}d\tau_{2}\
\int\limits_{\tau_{0}}^{\tau_{2}}d\tau_{1}\
{\cal{O}}_{1}(x+\xi(\tau_{1}))\star
{\cal{O}}_{2}(x+\xi(\tau_{2}))+\int\limits_{\tau_{0}}^{\tau}d\tau_{1}\
\int\limits_{\tau_{0}}^{\tau_{1}}d\tau_{2}\
{\cal{O}}_{2}(x+\xi(\tau_{2}))\star
{\cal{O}}_{1}(x+\xi(\tau_{1})),
\end{eqnarray*}
with $\xi(\tau)\equiv \Theta^{\mu\nu}p_{\nu}\tau$. In the momentum
space, after integrating over $y$, we get
\begin{eqnarray}\label{NS53}
\lefteqn{\hspace{-0.5cm}
{\cal{A}}_{inv.}^{nonplanar}(x)\Bigg|_{first\
term}=\lim\limits_{M\to \infty}+\frac{ig^{2}}{M^{4}}\int
\frac{d^{4}k}{(2\pi)^{4}}\ e^{+\frac{k^{2}}{M^{2}}}\int
\frac{d^{4}k_{1}}{(2\pi)^{4}}\frac{d^{4}k_{2}}{(2\pi)^{4}}\
e^{-i(k_{1}+k_{2})x}\ e^{2i(k_{1}+k_{2})\times k}
}\nonumber\\
&&\hspace{-1cm}\times
\bigg\{\int\limits_{0}^{1}d\tau_{2}\int\limits_{0}^{\tau_{2}}d\tau_{1}\
e^{-ik_{1}\times
k_{2}(1+2\tau_{1}-2\tau_{2})}+\int\limits_{0}^{1}d\tau_{1}\int\limits_{0}^{\tau_{1}}d\tau_{2}\
e^{+ik_{1}\times k_{2}(1-2\tau_{1}+2\tau_{2})} \bigg\}\
F_{\mu\nu}(k_{1})\ \tilde{F}^{\mu\nu}(k_{2}),
\end{eqnarray}
with $k_{1}\times
k_{2}\equiv\frac{\Theta^{\mu\nu}}{2}k_{1\mu}k_{2\nu}$. Performing
the integration over $\tau_{i}$, $i=1,2$ and  using
\begin{eqnarray}\label{NS54}
\int\limits_{0}^{1}d\tau_{2}\int\limits_{0}^{\tau_{2}}d\tau_{1}\
e^{-ik_{1}\times
k_{2}(1+2\tau_{1}-2\tau_{2})}+\int\limits_{0}^{1}d\tau_{1}\int\limits_{0}^{\tau_{1}}d\tau_{2}\
e^{+ik_{1}\times k_{2}(1-2\tau_{1}+2\tau_{2})}
=\frac{\sin(k_{1}\times k_{2})}{k_{1}\times k_{2}},
\end{eqnarray}
we obtain
\begin{eqnarray}\label{NS55}
{\cal{A}}_{inv.}^{nonplanar}(x)\Bigg|_{first\ term}&=&\lim\limits_{M\to \infty}
+\frac{ig^{2}}{M^{4}}\int \frac{d^{4}k}{(2\pi)^{4}}\ e^{+\frac{k^{2}}{M^{2}}}\int \frac{d^{4}k_{1}}{(2\pi)^{4}}
\frac{d^{4}k_{2}}{(2\pi)^{4}}\ e^{-i(k_{1}+k_{2})x}
\nonumber\\
&&\hspace{1cm}\times e^{2i(k_{1}+k_{2})\times k}\ F_{\mu\nu}(k_{1})\frac{\sin(k_{1}\times k_{2})}{k_{1}\times k_{2}}
\tilde{F}^{\mu\nu}(k_{2}).
\end{eqnarray}
Next the integration over $k$ in the Euclidean space can be performed using
\begin{eqnarray*}
e^{+\frac{k^{2}}{M^{2}}}\ e^{2i(k_{1}+k_{2})\times k}=e^{+\frac{1}{M^{2}}(k+\frac{i}{2}M^{2} (k_{1}+k_{2})^{\mu}
\Theta_{\mu\nu})^{2}}\ e^{-\frac{M^{2}}{4}(k_{1}+k_{2})\circ (k_{1}+k_{2})},
\end{eqnarray*}
with $q\circ q\equiv q_{\mu}\Theta^{\mu\nu}\Theta_{\nu\rho}q^{\rho}$. The first term of
${\cal{A}}_{inv.}^{nonplanar}$ is therefore
\begin{eqnarray}\label{NS56}
\lefteqn{
{\cal{A}}_{inv.}^{nonplanar}(x)\Bigg|_{first\ term}=\lim\limits_{M\to \infty}-\frac{g^{2}}{16\pi^{2}}
\int\frac{d^{4}k_{1}}{(2\pi)^{4}}\ \frac{d^{4}k_{2}}{(2\pi)^{4}}\ e^{-\frac{M^{2}}{4}(k_{1}+k_{2})\circ (k_{1}+k_{2})}
}\nonumber\\
&&\hspace{4cm}\times\ e^{-ik_{1}x} F_{\mu\nu}(k_{1})\ \frac{\sin(k_{1}\times k_{2})}{k_{1}\times k_{2}}\
\tilde{F}^{\mu\nu}(k_{2})\ e^{-ik_{2}x}.
\end{eqnarray}
Now we are in the position to discuss the celebrated UV/IR mixing
introduced in \cite{minwalla}. As it is explained there, in
noncommutative gauge theories, involving both the UV cutoff $M$
and the IR cutoff $|\Theta q|$, the two limits $M\to \infty$ and
$|\Theta q|\to 0$ do not commute. To show this in the case of
anomalies, let us define $q\equiv k_{1}+k_{2}$ in (\ref{NS56}),
and consider the limit $\frac{M^{2}\ q\circ q}{4}\gg 1$ or
$\frac{q\circ q}{4}\gg\frac{1}{M^{2}}$. This limit is equivalent
with taking first the limit $M\to \infty$ and then $|\Theta q|\to
0$. In this case, even before taking $|\Theta q|\to 0$ the
exponent $\exp(-\frac{M^{2}q\circ q}{4})$ vanishes. Thus in limit
$M\to \infty$ and for any value of $|\Theta q|$, the first term in
the expansion of ${\cal{A}}_{inv.}^{nonplanar}$ vanishes. In the
opposite case, {\it i.e.} when we take first $|\Theta q|\to 0$ and
then $M\to \infty$, a finite anomaly arises. This limit can be
understood as the limit $\frac{M^{2}\ q\circ q}{4}\ll 1$ or
$\frac{q\circ q}{4}\ll\frac{1}{M^{2}}$, too \cite{minwalla}. In
this case the exponent $\exp(-\frac{M^{2}q\circ q}{4})\to 1$ and
we are left with a finite nonplanar anomaly.
\begin{eqnarray}\label{NS57}
{\cal{A}}_{inv.}^{nonplanar}(x)\Bigg|_{first\
term}=-\frac{g^{2}}{16\pi^{2}}\ F_{\mu\nu}(x)
\star'\tilde{F}^{\mu\nu}(x),
\end{eqnarray}
where the field strength tensors $F_{\mu\nu}$'s are as before
defined in the noncommutative space by (\ref{NS4}). Taking the
limit $M\to \infty$, which is performed in both cases, remove the
unphysical cutoff dependence in the final expression for the
anomaly.
\par
Summarizing the above results from UV/IR mixing, the invariant
(nonplanar) anomaly vanishes by taking the limit $M\to \infty$
keeping $|\Theta q|$ arbitrary but finite. But, when we take the
$|\Theta q|\to 0$ limit before $M\to \infty$, we find
\begin{eqnarray}\label{NS58}
{\cal{A}}_{inv.}^{nonplanar}(x)=
-\frac{g^{2}}{16\pi^{2}}F_{\mu\nu}(x)
\star'\tilde{F}^{\mu\nu}(x)+\cdots,
\end{eqnarray}
where the ellipses denote the higher order terms arising from the
expansion of the Wilson line in the orders of the external gauge
field. Note that the contributions from the Wilson line guarantees
the $\star$-gauge invariance of the unintegrated form of the
anomaly. This is shown in \cite{banerjee}, where the unintegrated
invariant anomaly is determined using a Seiberg-Witten map. One of
the conditions which must be fullfilled to find this map is that
the integrated form of the invariant anomaly should be identical
with the integrated form of the $\star$-gauge covariant anomaly
(\ref{NS24}).

Indeed, integrating the expression (\ref{NS58}) over the
noncommutative space, leads to a gauge invariant result identical
with the integrated form of the covariant anomaly (\ref{NS24}). If
the divergence of $j_{inv.}^{\mu,5}$ did vanish for all values of
$|\Theta q|$ \cite{intriligator}, a contradiction would arise when
we looked at the integrated version of the covariant and invariant
anomaly. This contradiction and its resolution, using the
nonplanar anomaly \cite{ned2}, was first discussed in
\cite{armoni}, where the non-conservation of the axial charge
corresponding to two currents $J_{cov.}^{\mu,5}$ and
$j_{inv.}^{\mu,5}$ is studied. Using the cyclic symmetry of the
$\star$-product under integration, both currents
$J_{cov.}^{\mu,5}$ and $j_{inv.}^{\mu,5}$ lead to the same gauge
invariant axial charge defined by $Q_{5}\equiv \int d\hat{x}\
J_{cov.}^{0,5}=\int d\hat{x}\ j_{inv.}^{0,5}$. The covariant
anomaly (\ref{NS24}) implies that $Q_{5}$ is not conserved. Hence
a vanishing of $j_{inv.}^{\mu,5}$ for all value of $|\Theta q|$
would lead to an inconsistency. However, this obvious
contradiction is resolved, if we note that a finite (nonplanar)
anomaly arises from a singularity at $|\Theta q|\to 0$ as is given
in (\ref{NS58}). The integrated version of the nonplanar anomaly
receives therefore a finite contribution from $|\Theta q|\to 0$
limit.
\par
As we have explained above, the Wilson line in the final
expression for the invariant anomaly is necessary to preserve the
$\star$-gauge invariance of the unintegrated form of the anomaly.
Expanding the Wilson line in the order of the external gauge field
leads to infinitely many terms, which can be understood as the
contribution from infinitely many diagrams, where more and more
gauge fields are inserted to the ordinary triangle diagrams. As it
is stated in \cite{ned2}, this result can be regarded as the
noncommutative generalization of Adler-Bardeen's
nonrenormalization theorem \cite{adler2}. Note that all additional
terms due to the Wilson line attachment, and higher order in the
external gauge field can be written as a total derivative. Thus,
once we integrate over the full expression of the invariant
(nonplanar) anomaly including the Wilson line, we are left with
the same integrated form of the covariant (planar) anomaly. This
question is also addressed in \cite{banerjee}.
\par
We shall further note that the cancellation of the nonplanar
anomaly in the case when we take first $M\to \infty$ and for
arbitrary but finite $|\Theta q|$, can be understood in the
framework of a noncommutative Green-Schwarz mechanism of anomaly
cancellation \cite{intriligator,armoni}. In fact, in the this
limit the gauge theory is still coupled to the string theory, so
that a nonvanishing tree level contribution from Ramond-Ramond
charges is responsible for the cancellation of the nonplanar
anomaly in a mechanism similar to the Green-Schwarz mechanism of
anomaly cancellation in the commutative case \cite{green}.
\subsubsection*{3\hspace{0.3cm}Noncommutative Konishi Anomaly}\label{Sec3}

\setcounter{section}{3} \setcounter{equation}{0} The action of
noncommutative ${\cal{N}}=1$ supersymmetric $U(1)$ gauge theory is
given by
\begin{nedalph}\label{NS3-1a}
S=S_{matter}+S_{gauge},
\end{eqnarray}
with
\begin{eqnarray}\label{NS3-1b}
S_{matter}=\int d^{4}x\ d^{2}\theta\ d^{2}\bar{\theta}\ \bar{\Phi}(x,\bar{\theta})\star
e^{V(x,\theta,\bar{\theta})}\star \Phi(x,\theta),
\end{eqnarray}
where $\Phi$ and $V$ are the standard chiral matter field and gauge field supermultiplets, and
\begin{eqnarray}\label{NS3-1c}
S_{gauge}=-\frac{1}{16\pi i}\int d^{4}x\ d^{2}\theta\ \tau
W_{\alpha}(x,\theta)\star W^{\alpha}(x,\theta)+\mbox{h.c.},
\end{nedalph}
\par\noindent
with holomorphic coupling $\tau\equiv \frac{4\pi
i}{g^{2}}+\frac{\vartheta}{2\pi}$ including the gauge coupling and
the $\vartheta$-angle. Using the definition
$W_{\alpha}=\bar{D}^{2}(e^{-V})_{\star}\
D_{\alpha}(e^{V})_{\star}$, it can be shown that the gauge action
involves both $F_{\mu\nu}\star F_{\mu\nu}$ and $F_{\mu\nu}\star
\tilde{F}^{\mu\nu}=\varepsilon_{\mu\nu\lambda\rho}\
F^{\mu\nu}\star F^{\lambda\rho}$ term that are proportional to
$\frac{1}{g^{2}}$ and the $\vartheta$-angle, respectively. The
above action (\ref{NS3-1a}-c) is invariant under local
$\star$-gauge transformation of matter fields in the fundamental
representation
\begin{eqnarray}\label{NS3-3}
\Phi(x,\theta)\to \Phi'(x,\theta)=
e^{i\Lambda(x,\theta)}\star \Phi(x,\theta),\hspace{1cm}\mbox{and}
\hspace{1cm}\bar{\Phi}(x,\bar{\theta})\to \bar{\Phi'}=\bar{\Phi}
(x,\bar{\theta})\star e^{-i\bar{\Lambda}(x,\bar{\theta})},
\end{eqnarray}
and $\star$-gauge transformation of the gauge supermultiplet
\begin{eqnarray}\label{NS3-4}
e^{V}\to e^{i\bar{\Lambda}}\star e^{V}\star e^{-i\Lambda},
\end{eqnarray}
where $\Lambda$ and $\bar{\Lambda}$ are arbitrary chiral and antichiral superfields.
\par
Consider now the matter part of the action (\ref{NS3-1b}). It is
easy to check the invariance of this action under global
transformation of the chiral matter field $\Phi$
\begin{eqnarray}\label{NS3-5}
\Phi\to \Phi'=e^{iA}\Phi,
\end{eqnarray}
where $A$ is a constant parameter. To find the global currents of
this theory, we proceed as in the nonsupersymmetric theory and
define two independent local change of variables corresponding to
the same global transformation (\ref{NS3-5})
\begin{eqnarray}\label{NS3-6}
\delta_{A}\Phi(z)&=&iA(z)\star\Phi(z),\nonumber\\
\delta_{A}\Phi(z)&=&i\Phi(z)\star A(z).
\end{eqnarray}
Here $z$ stands for the collective supercoordinates $x$ and
$\theta$. The corresponding currents to this change of variables
are
\begin{eqnarray}\label{NS3-7}
J_{cov.}(z)&\equiv &\Phi(z)\star\bar{\Phi}(z)\star e^{V(z)},\nonumber\\
J_{inv.}(z)&\equiv& \bar{\Phi}(z)\star e^{V(z)}\star\Phi(z).
\end{eqnarray}
The first and second currents are covariant and invariant under
the $\star$-gauge transformation, respectively. In the following
we will determine the anomaly in the above symmetry corresponding
to the currents (\ref{NS3-7}) separately.
\subsubsection*{3.1\hspace{0.3cm}Covariant (Planar) Konishi Anomaly}\label{Sec31}
Using the method presented in \cite{konishi}, we will calculate
the anomaly corresponding to the covariant current $J_{cov.}$
(\ref{NS3-7}).  We consider the invariance of the partition
function of noncommutative ${\cal{N}}=1$ supersymmetric $U(1)$
\begin{eqnarray}\label{NS3-8}
{\cal{Z}}=\int {\cal{D}}\Phi\ {\cal{D}}\bar{\Phi}\ e^{-iS_{matter}},
\end{eqnarray}
under the local {\it fundamental} change of variable
$\delta_{A}\Phi(z)=iA(z) \star\Phi(z)$. The variation of the
matter field action $S_{matter}$ corresponding to this change of
variable is given by
\begin{eqnarray}\label{NS3-9}
\delta_{A} S_{matter}=\int d^{8}z\ A(z)\star J_{cov.}(z).
\end{eqnarray}
The Jacobian ${\cal{J}}$ of this transformation can be easily
calculated using the same  method from nonsupersymmetric field
theory presented in the previous section, and reads
\begin{eqnarray}\label{NS3-10}
{\cal{J}}=\exp\left(i\sum\limits_{n}\int d^{8}z\ A(z)\star\Phi_{n}(z)\star\bar{\Phi}_{n}(z)\right),
\end{eqnarray}
where $\phi_{n}(z)$'s are a complete orthonormal basis of the
Hilbert space. Combining the two results (\ref{NS3-9}) and
(\ref{NS3-10}), and using the invariance of the  partition
function under the above {\it fundamental} change of variables, we
arrive first at
\begin{eqnarray}\label{NS3-11}
-\frac{\bar{D}^{2}}{4}J_{cov.}(z)=-i\sum\limits_{n}\frac{\bar{D}^{2}}{4}\left(\Phi_{n}(z)\star\bar{\Phi}_{n}(z)\right).
\end{eqnarray}
The r.h.s. of the above equation is then to be regulated using a Gaussian damping factor $e^{L/M^{2}}$,
with the operator
\begin{eqnarray}\label{NS3-12}
L\equiv \frac{1}{16}\bar{D}^{2}(e^{-V})_{\star}\ D^{2}(e^{V})_{\star},
\end{eqnarray}
and $M$ the Pauli-Villars mass. The chiral operator $L$ is
manifestly supersymmetric invariant, gauge covariant and contains
$\DS$ as the lowest component. Since the damping factor transforms
covariantly under $\star$-gauge transformation, we have to insert
it on the r.h.s. of (\ref{NS3-11}) so that the resulting
expression remains gauge covariant. We therefore obtain
\begin{eqnarray}\label{NS3-13}
-\frac{\bar{D}^{2}}{4}J_{cov.}(z)=\lim\limits_{M\to \infty}-i\sum\limits_{n}\ e^{\frac{L}{M^{2}}}
\star \frac{\bar{D}^{2}}{4}\left(\Phi_{n}(z)
\star\bar{\Phi}_{n}(z)\right).
\end{eqnarray}
Going now to the Fourier space we get
\begin{eqnarray}\label{NS3-14}
-\frac{\bar{D}^{2}}{4}J_{cov.}(z)&=&\lim\limits_{M\to\infty}\ -i
\int \frac{d^{4}k_{1}}{(2\pi)^{4}}\ e^{\frac{L}{M^{2}}}\star\frac{\bar{D}^{2}}{4}\ e^{ik_{1}z}
\star e^{-ik_{2}z}\sum\limits_{n}\Phi_{n}(k_{1})
\bar{\Phi}_{n}(k_{2}),\nonumber\\
&=&\lim\limits_{M\to \infty}-\frac{i}{2M^{4}}\int
\frac{d^{4}k}{(2\pi)^{4}}\ e^{+\frac{k^{2}}{M^{2}}}\
W_{\alpha}(x,\theta)\star W^{\alpha}(x,\theta),
\end{eqnarray}
integrating over $k$ we arrive at the planar (covariant) Konishi anomaly corresponding to $J_{cov.}$
\begin{eqnarray}\label{NS3-15}
-\frac{\bar{D}^{2}}{4}\ J_{cov.}(z)={\cal{S}}_{cov.}^{planar}(z),\hspace{1cm}\mbox{with}\hspace{1cm}
{\cal{S}}_{cov.}^{planar}(z)\equiv-\frac{1}{32\pi^{2}}\ W_{\alpha}(z)\star W^{\alpha}(z).
\end{eqnarray}
This is the only finite term surviving the limit $M\to \infty$ (see also \cite{konishi}).
As $W_{\alpha}$'s transform covariantly under $\star$-gauge transformation
\begin{eqnarray}\label{NS3-16}
W_{\alpha}(z)\to e^{i\Lambda(z)}\star W_{\alpha}(z)\star e^{-i\Lambda(z)},
\end{eqnarray}
clearly ${\cal{S}}_{cov.}^{planar}(z)$ transforms also
covariantly. The full Konishi equation for nonvanishing tree level
superpotential reads
\begin{eqnarray}\label{kpl}
-\frac{\bar{D}^{2}}{4}\ J_{cov.}(z)=\Phi(z)\star\frac{\partial W_{tree}(z)}{\partial \Phi(z)}+
{\cal{S}}_{cov.}^{planar}(z).
\end{eqnarray}
\subsubsection*{3.2\hspace{0.3cm}Invariant (Nonplanar) Konishi Anomaly}\label{Sec32}
In this section we will derive the Konishi anomaly corresponding to the invariant current
\begin{eqnarray*}
J_{inv.}(z)=\bar{\Phi}(z)\star e^{V}\star \Phi(z),
\end{eqnarray*}
of (\ref{NS3-7}). The  {\it antifundamental} change of variable $\delta_{A}\Phi(z)=i\Phi(z)\star A(z)$ leads to
\begin{eqnarray}\label{NS3-21}
\delta_{A}S_{matter}=\int d^{8}z\ A(z)\star J_{inv.}(z),
\end{eqnarray}
and the Jacobian reads
\begin{eqnarray}\label{NS3-22}
{\cal{J}}=\exp\left(i\sum\limits_{n}\int d^{8}z\ A(z)\star\bar{\Phi}_{n}(z)\star\Phi_{n}(z)\right),
\end{eqnarray}
with $\Phi_{n}(z)$ the same orthonormal and complete basis as was introduced previously.
Combining now the results from (\ref{NS3-21}) and (\ref{NS3-22}), and using the invariance of the
partition function under global transformation $\delta_{A}\Phi=iA\Phi$ we arrive first at
\begin{eqnarray}\label{NS3-23}
-\frac{\bar{D}^{2}}{4}\ J_{inv.}=-i\sum\limits_{n}\frac{\bar{D}^{2}}{4}\left(\bar{\Phi}_{n}(z)\star\Phi_{n}(z)\right).
\end{eqnarray}
To regulate the r.h.s. we have to insert a  $\star$-gauge
invariant version of gauge covariant damping factor $e^{L/M^{2}}$.
As in the nonsupersymmetric case we smear the operator $L$
(\ref{NS3-12}) along an open Wilson line ${\cal{W}}(y,\theta;C)$
\begin{eqnarray}\label{NS3-24}
-\frac{\bar{D}^{2}}{4} J_{inv.}(x,\theta)&=&\lim\limits_{M\to\infty} -i \int\frac{d^{4}p}{(2\pi)^{4}}\
e^{-ipx}\ \int d^{4}y\ \sum\limits_{n}\frac{\bar{D}^{2}}{4}\ P_{\star}\bigg[\bar{\Phi}_{n}(y,\theta)\star
{\cal{W}}\left(y,\theta;C\right)
\nonumber\\
&&\star \exp\left(\frac{1}{M^{2}}\int\limits_{0}^{1}d\tau\ L(x+\tilde{p}\tau,\theta)\right)\star
\Phi_{n}(y,\theta)\bigg]\star e^{ipy}.
\end{eqnarray}
Here we have introduced the supersymmetric version of an open
Wilson line ${\cal{W}}(y,\theta;C)$, as was used for the
nonsupersymmetric case (\ref{x3}). The explicit form of the
$\star$-generalization of the
  supersymmetric Wilson line for the commutative case
introduced in \cite{mansouri} reads
\begin{eqnarray}\label{NS3-16a}
{\cal{W}}(C_{z_{1},z_{2}})=\exp\left(\int\limits_{C_{z_{1},z_{2}}} ds\ \dot{z}^{A}\ A_{A}\right)_{\star},
\hspace{1cm}\dot{z}^{A}=\frac{dz^{M}}{ds}\ e_{M}^{A},
\end{eqnarray}
with $e_{M}^{A}$ the vierbein matrix \cite{wess}
\begin{eqnarray}\label{NS3-uu}
e_{A}^{M}\equiv\left(
  \begin{array}{ccc}
\delta_{\alpha}^{m}&0&0\\
i\sigma_{\alpha\dot{\beta}}^{m}\bar{\theta}^{\dot{\beta}}&\delta_{\alpha}^{\mu}&0\\
i\left(\theta\sigma\epsilon\right)^{\dot{\alpha}}&0&\delta_{\dot{\mu}}^{\dot{\alpha}}\\
\end{array}
\right).
\end{eqnarray}
We have used the following notations
\begin{eqnarray}\label{NS3-16b}
z^{M}=(x^{m}, \theta^{\mu}, \bar{\theta}_{\dot{\mu}}), &\hspace{1cm}&e^{M}_{A}\frac{\partial}{\partial z^{M}}
\equiv D_{A}\equiv \left(\partial_{\alpha}, D_{\alpha}, \bar{D}^{\dot{\alpha}}\right),\nonumber\\
D_{\alpha}=\frac{\partial}{\partial\theta^{\alpha}}+i\left(\sigma^{m}\bar{\theta}\right)_{\alpha}
\partial_{m},&\hspace{1cm}&\bar{D}^{\dot{\alpha}}=\frac{\partial}{\partial\bar{\theta}^{\dot{\alpha}}}+
i\left(\theta\sigma^{m}\epsilon\right)^{\dot{\alpha}}\partial_{m},
\end{eqnarray}
and introduced the superfield $A_{A}$ for the gauge field supermultiplets $U$ and $V$ (matrices in the Lie algebra)
\begin{eqnarray}\label{NS3-16c}
A_{\alpha}\equiv e^{-V}\ D_{\alpha}\ e^{V},\hspace{1cm}A_{\dot{\alpha}}\equiv e^{-U}\ \bar{D}_{\dot{\alpha}}\
e^{U},\nonumber\\
A_{A}\equiv \frac{1}{4}i\bar{\sigma}_{A}^{\dot{\beta}\alpha}\left(D_{\alpha}A_{\dot{\beta}}+\bar{D}_{\dot{\beta}}
A_{\alpha}+\{A_{\alpha},A_{\dot{\beta}}\}\right).
\end{eqnarray}
As in the nonsupersymmetric case the supersymmetric Wilson line (\ref{NS3-16a}) transforms under
$\star$-gauge transformation as
\begin{eqnarray}\label{NS3-17}
{\cal{W}}(x,\theta;\ell)\to e^{i\Lambda(x,\theta)}\star {\cal{W}}(x,\theta;\ell)\star e^{-i\Lambda(x+\ell,\theta)}.
\end{eqnarray}
Consider again the expression on the r.h.s. of (\ref{NS3-24}).
Going to the Fourier space and performing the same manipulations as in equation (\ref{NS51a}-b) we arrive at
\begin{nedalph}\label{NS3-26a}
-\frac{\bar{D}^{2}}{4}J_{inv.}=S_{inv.}^{nonplanar},
\end{eqnarray}
where
\begin{eqnarray}\label{NS3-26b}
\lefteqn{\hspace{-1.5cm}S_{inv.}^{nonplanar}(x,\theta)\equiv\lim\limits_{M\to
\infty} -\frac{i}{M^{4}}\int \frac{d^{4}p}{(2\pi^{4})}\
e^{-ipx}\int \frac{d^{4}k}{(2\pi)^{4}}e^{+\frac{k^{2}}{M^{2}}}\
\int d^{4}y\
P_{\star}\bigg[{\cal{W}}(y-\tilde{k},\theta;\ell=\tilde{p})
}\nonumber\\
&&\times   \int\limits_{0}^{1} d\tau_{1}\
W_{\alpha}(y-\tilde{k}+\tau_{1}\tilde{p},\theta)\star\int\limits_{0}^{1}
d\tau_{2}\
W^{\alpha}(y-\tilde{k}+\tau_{2}\tilde{p},\theta)\bigg]\star
e^{ipy}.
\end{nedalph}
\par\noindent
Using now the expansion of the supersymmetric Wilson line
(\ref{NS3-16a}) and going through the same algebraic manipulations
following  from (\ref{NS51a}-b) to  (\ref{NS56}) in the
nonsupersymmetric case, we obtain the lowest term of the nonplanar
Konishi anomaly in the expansion of the supersymmetric Wilson line
in powers of the superfield $A_{A}$
\begin{eqnarray}\label{NS3-27}
\lefteqn{S_{inv.}^{nonplanar}(x,\theta)\Bigg|_{first\ term}=-\frac{1}{32\pi^{2}}\int
\frac{d^{4}k_{1}}{(2\pi)^{4}}\ \frac{d^{4}k_{2}}{(2\pi)^{4}}
e^{-\frac{M^{2}}{4}\ q\circ q}
}\nonumber\\
&&\times
e^{-ik_{1}x}\ W_{\alpha}(k_{1},\theta)\ \frac{\sin\left(k_{1}\times k_{2}\right)}{(k_{1}\times k_{2})}\
W_{\alpha}(k_{2},\theta)\ e^{-ik_{2}x}+\mbox{higher order terms},
\end{eqnarray}
with $q\equiv k_{1}+k_{2}$. Here similar UV/IR mixing phenomena as
in the nonsupersymmetric case occur, {\it i.e.} the invariant
(nonplanar) Konishi anomaly appears only when we take first
$|\Theta q|to 0$ and then $M\to \infty$ [see the discussion
leading to (\ref{NS57}) in the previous section]
\begin{eqnarray}\label{NS3-28}
{\cal{S}}_{inv.}^{nonplanar}(x,\theta)\Bigg|_{first\
term}=-\frac{1}{32\pi^{2}}\ W_{\alpha}(x,\theta)\star'
W^{\alpha}(x,\theta).
\end{eqnarray}
In summary the invariant (nonplanar) Konishi anomaly vanishes if
we take first $M\to \infty$ keeping $|\Theta q|$ arbitrary but
finite and appears as a singularity at $|\Theta q|\to 0$
limit
\begin{eqnarray}\label{NS3-29}
{\cal{S}}_{inv.}^{nonplanar}(x,\theta)=
-\frac{1}{32\pi^{2}}W_{\alpha}(x,\theta) \star'
W^{\alpha}(x,\theta)+\cdots,
\end{eqnarray}
where the ellipses denote the contribution of the higher order
terms in the external gauge multiplet arising from the expansion
of the attached Wilson line.
\\
The full Konishi equation, including the contributions of the tree
level superpotential, is given by
\begin{eqnarray}\label{knpl}
-\frac{\bar{D}^{2}}{4} J_{inv.}(z)=\frac{\partial W_{tree}(z)}{\partial\Phi(z)}\star \Phi(z)+
{\cal{S}}_{inv.}^{nonplanar}(z).
\end{eqnarray}
\subsubsection*{4\hspace{0.3cm}Effective Superpotential for Noncommutative ${\cal{N}}=1$ Supersymmetric $U(1)$}
\label{Sec4}
\setcounter{section}{4} \setcounter{equation}{0}
In this section, we will determine the effective superpotential
 of noncommutative ${\cal{N}}=1$ supersymmetric $U(1)$ gauge
 theory.
Assuming that the original theory consists of one gauge
supermultiplet and $2N_{f}$ fundamental and antifundamental chiral
matter fields $Q_{i}$ and $\tilde{Q}_{i}$ with $i=1,\cdots,N_{f}$,
the effective field theory will depend on the noncommutative
generalization of the meson and gaugino bilinears. The fundamental
matter multiplet $Q_{i}$ and antifundamental matter multiplet
$\tilde{Q}_{i}$ transform as
\begin{eqnarray}\label{NS4-1}
Q_{i}\to e^{i\Lambda}\star
Q_{i},\hspace{1cm}\mbox{and}\hspace{1cm}\tilde{Q}_{i}\to \tilde{Q}_{i}\star e^{-i\Lambda},
\end{eqnarray}
and the noncommutative $\star$-gauge invariant meson is defined by
\begin{eqnarray}\label{NS4-2}
T_{ij}\equiv\tilde{Q}_{i}\star
Q_{j},\hspace{1cm}\mbox{with}\hspace{1cm}i,j=1,\cdots,N_{f}.
\end{eqnarray}
Further, $S\propto W_{\alpha}\star W^{\alpha}$ and $S'\propto
W_{\alpha}\star' W^{\alpha}+\cdots$, with the extra terms denoting
the contributions from the expansion of the attached Wilson line,
will be used as the relevant gaugino superfields.
\par
As in the ordinary supersymmetric commutative theories
\cite{seiberg}, it can be shown using holomorphy and other
symmetry arguments that the effective superpotential consists of a
tree level superpotential $W_{tree}$ and a nonperturbative term,
the Veneziano-Yankielowicz (VY) dynamical superpotential $W_{dyn}$
\cite{vy,vyt}
\begin{eqnarray}\label{NS4-4}
W_{eff.}(T, S;\Lambda_{N_{f}})=W_{tree}(T;m, \lambda)+W_{dyn}(T,S;\Lambda_{N_{f}}),
\end{eqnarray}
where $m$ and $\lambda$ are bare parameters of the
superpotential\footnote{A complete proof of non-renormalization
theorem of noncommutative supersymmetric theories will be
presented elsewhere.} and $\Lambda_{N_{f}}$ is  the holomorphic
intrinsic scale for a theory with $N_{f}$ massless flavors. It is
defined as in ordinary commutative supersymmetric theories by the
$\vartheta$-angle
\begin{eqnarray}
\Lambda_{N_{f}}=|\Lambda_{N_{f}}|\ e^{i\vartheta/\beta_{N_{f}}}.
\end{eqnarray}
Here $|\Lambda_{N_{f}}|$ is the intrinsic scale of the
noncommutative U(1) gauge theory that enters through dimensional
transmutation. It is given by the coefficient $\beta_{N_{f}}$ of
the one-loop $\beta$-function $\beta(g)=-\frac{ g^{3}}{16\pi^{2}}\
\beta_{N_{f}}$ of the theory
\begin{eqnarray}\label{NS4-4a}
|\Lambda_{N_{f}}|\equiv\mu
\exp\left(-\frac{8\pi^{2}}{\beta_{N_{f}} g^{2}(\mu)}\right).
\end{eqnarray}
The subscript $N_{f}$ denotes the number of massless flavors.

Before calculating the superpotential of noncommutative
supersymmetric $U(1)$ theory, we will compute the one-loop
$\beta$-function of the theory and also the anomaly corresponding
to $U_{R}(1)$ symmetry. These are necessary to determine the
effective superpotentials, as in the ordinary commutative
supersymmetric gauge theories, using the selection rules.
\subsubsection*{4.1\hspace{0.3cm}Prerequisites}
\subsubsection*{One-Loop $\beta$-Function}
As we will need the $\beta$-function of the noncommutative theory
for our calculations, we will present a supersymmetric extension
of the result in \cite{khoze} and \cite{alvarez}, where the effect
of nonplanar diagrams on the one-loop $\beta$-function of the
noncommutative nonsupersymmetric $U(1)$ are studied.
\par
Let us consider a noncommutative theory consisting of $n_{f}$ Weyl
fermions in the fundamental representation and $n_{s}$ complex
scalars. Using the background field method, the effective coupling
of the theory is given by
\begin{eqnarray}
\frac{1}{g^{2}(p)}=\frac{1}{g^{2}}+\Pi(p),
\end{eqnarray}
with $\Pi(p)$ is defined by the vacuum polarization tensor
\begin{eqnarray}
\Pi_{\mu\nu}=(p^{2}\delta_{\mu\nu}-p_{\mu}p_{\nu})\Pi(p)+\frac{\tilde{p}_{\mu}\tilde{p}_{\nu}}{\tilde{p}^{2}}\
\tilde{\Pi}(p).
\end{eqnarray}
It is given explicitly by
\begin{eqnarray}\label{non}
\Pi(p)&=&\frac{1}{8\pi^{2}}\Bigg[\int\limits_{0}^{1}dx\
\sum\limits_{j_{a}}\alpha(j_{a})\left(d(j_{a})(1-2x)^{2}-4C(j_{a})\right)
\left(\ln\frac{\Delta}{4\pi\mu^{2}}+2K_{0}\left(|\tilde{p}|\sqrt{\Delta}\right)\right)\nonumber\\
&&\hspace{1cm}+\frac{1}{2}\int\limits_{0}^{1}dx\
\sum\limits_{j_{f}}\alpha(j_{f})\left(d(j_{f})(1-2x)^{2}-4C(j_{f})\right)\
\ln\frac{\Delta}{4\pi\mu^{2}}\bigg],
\end{eqnarray}
with $\Delta=x(1-x)p^{2}$ and the modified Bessel function
\begin{eqnarray*}
K_{\nu}(\alpha z)\equiv
\frac{\alpha^{\nu}}{2}\int\limits_{0}^{\infty}\frac{dt}{t^{\nu+1}}
e^{-\frac{z}{2}(t+\frac{\alpha^{2}}{t})},
\end{eqnarray*}
which appears only in the contribution of the nonplanar diagrams.
Adding over all $j_{a}$ and $j_{f}$ for the fields in adjoint
(ghosts and gauge fields) and fundamental (Weyl fermions and
complex scalars) representation, respectively, we arrive first at
\begin{eqnarray}\label{m1}
\Pi(p)&=&\frac{1}{8\pi^{2}}\int
\limits_{0}^{1}dx\bigg[(4-(1-2x)^{2})\left(\ln\frac{\Delta}{4\pi\mu^{2}}+
2K_{0}(|\tilde{p}|\sqrt{\Delta})\right)
\nonumber\\
&&\hspace{1.5cm}-\left(\frac{n_{f}}{2}(1-(1-2x)^{2})+
\frac{n_{s}}{2}(1-2x)^{2}\right)\ln\frac{\Delta}{4\pi\mu^{2}}
\bigg].
\end{eqnarray}
This expression can be compared with the result in \cite{alvarez}.
The values of $\alpha(j), d(j)$ and $C(j)$ are listed below
\begin{center}
\begin{tabular}{|c||c|c||c||c|c|}
\hline
&ghosts&gauge fields&&Weyl fermions&Complex scalars\\
\hline \hline
$d(j_{a})$&1&4&$d(j_{f})$&2&1 \\
$C(j_{a})$&0&2&$C(j_{f})$&$\frac{1}{2}$&0\\
$\alpha(j_{a})$&1&$-\frac{1}{2}$&$\alpha(j_{f})$&$+\frac{n_{f}}{2}$&$-n_{s}$\\
\hline
\end{tabular}
\end{center}
Using
\begin{eqnarray}\label{NS4-10d}
\beta(g;p,\Theta)\equiv p_{\mu}\frac{\partial}{\partial
p_{\mu}}g(p)\cong -\frac{g^{3}}{2}\ p_{\mu}\frac{\partial \Pi
(p)}{\partial p_{\mu}},
\end{eqnarray}
and the relations
\begin{eqnarray}
p_{\mu}\frac{\partial}{\partial p_{\mu}}\
\ln\frac{\Delta}{4\pi\mu^{2}}=2,\hspace{1cm}\mbox{and}\hspace{1cm}
p_{\mu}\frac{\partial}{\partial p_{\mu}}
2K_{0}(|\tilde{p}|\sqrt{\Delta})=-4K_{1}(|\tilde{p}|\sqrt{\Delta})\
\frac{\tilde{p}^{2}\Delta}{|\tilde{p}|\sqrt{\Delta}},
\end{eqnarray}
the one-loop $\beta$-function of the theory for an arbitrary
noncommutativity parameter reads
\begin{eqnarray}\label{NS4-xx}
\beta(g;p,\Theta)\equiv -
\frac{g^{3}}{8\pi^{2}}\Bigg[-\frac{n_{f}}{3}-\frac{n_{s}}{6}+\int\limits_{0}^{1}dx\
(4-(1-2x)^{2}) \left(1-2K_{1}(|\tilde{p}|\sqrt{\Delta})
\frac{\tilde{p}^{2}\Delta}{|\tilde{p}|\sqrt{\Delta}}\right)\Bigg].
\end{eqnarray}
Note that here, the $\beta$-function depends in general on the
momentum $p$. This is due to the breaking of Lorentz invariance.
Momentum dependent $\beta$-functions were previously calculated
explicitly in the so called noncommutative dipole theories
\cite{soroush}.
\par
In the limit $|p|\gg\frac{1}{\sqrt{\Theta}}$,  only the planar
diagrams contribute to the $\beta$-function. This can be shown
using the relation $\lim\limits_{z\to\infty}K_{1}(z)=0$ in
(\ref{NS4-xx}). However, in the limit
$|p|\ll\frac{1}{\sqrt{\Theta}}$, both planar and nonplanar
diagrams contribute to the one-loop $\beta$-function. This can be
shown by taking the limit $\lim\limits_{z\to
0}K_{1}(z)=\frac{1}{z}$ in (\ref{NS4-xx}). This reflects the UV/IR
duality indicated in \cite{khoze} and \cite{alvarez}. For a
nonsupersymmetric theory with $n_{f}$ massless Weyl fermions in
the fundamental representation and $n_{s}$ complex scalars, the
$\beta$-function reads therefore
\begin{eqnarray}\label{beta3}
\beta(g;p,\Theta)=\left\{
\begin{array}{cccc}
-\frac{g^{3}}{16\pi^{2}}\beta_{\ell,n{f}}\qquad \mbox{with}&\beta_{\ell,n_{f}}=+2\left(\frac{11}{3}-\frac{n_{f}}{3}-\frac{n_{s}}{6}\right),&&|p|\gg\frac{1}{\sqrt{\Theta}},\\
-\frac{g^{3}}{16\pi^{2}}\beta_{s,n_{f}}\qquad
\mbox{with}&\beta_{s,n_{f}}=-2\left(\frac{11}{3}+\frac{n_{f}}{3}+\frac{n_{s}}{6}\right)
,&&|p|\ll\frac{1}{\sqrt{\Theta}}.\\
\end{array}
\right.
\end{eqnarray}
The subscript $\ell$ and $s$ in $\beta_{\ell,n_{f}}$ and
$\beta_{s,n_{f}}$ label the coefficient of the one-loop
$\beta$-function for two cases of $|p|\gg\frac{1}{\sqrt{\Theta}}$
and $|p|\ll\frac{1}{\sqrt{\Theta}}$, respectively. Note that since
the matter fields are in the fundamental representation, they do
not receive any contribution from nonplanar diagrams
\cite{khoze,alvarez}. The sign of the terms proportional to
$n_{f}$ and $n_{s}$ are therefore the same for
$|p|\gg\frac{1}{\sqrt{\Theta}}$ and
$|p|\ll\frac{1}{\sqrt{\Theta}}$ limits. As for the part arising
from the fields in the adjoint representation, {\it i.e.} terms
proportional to $11/3$ in (\ref{beta3}), they have opposite signs
on the first and second line of (\ref{beta3}). This is due to the
contribution from the nonplanar parts [see also \cite{alvarez},
where the same phenomenon occurs\footnote{Note that in
\cite{alvarez} since the matter fields are in the adjoint
representation, in contrast to our calculation, the terms
proportional to $n_{f}$ and $n_{s}$ have also the opposite signs
in $|p|\gg\frac{1}{\sqrt{\Theta}}$ and
$|p|\ll\frac{1}{\sqrt{\Theta}}$ limits.}]. According to this
result, for $|p|\ll \frac{1}{\sqrt{\Theta}}$ limit, the
coefficient of the one-loop $\beta$-function, including the minus
sign in front of $\frac{g^{3}}{16\pi^{2}}$, is positive. Thus in
this limit, the theory turns out to be IR free for all values of
$n_{f}$ and $n_{s}$.
\par
To arrive at the supersymmetric extension of one-loop
$\beta$-function in our case, we set $n_{f}=n_{s}$ and use the
values of $\alpha(j), d(j)$ and $C(j)$ from the table
\par
\begin{center}
\begin{tabular}{|c||c|c|c||c||c|c|}
\hline
&ghosts&gauge fields&Weyl gaugino&&Weyl fermions&Complex scalars\\
\hline \hline
$d(j_{a})$&1&4&2&$d(j_{f})$&2&1 \\
$C(j_{a})$&0&2&$+\frac{1}{2}$&$C(j_{f})$&$\frac{1}{2}$&0\\
$\alpha(j_{a})$&1&$-\frac{1}{2}$&$+\frac{1}{2}$&$\alpha(j_{f})$&$+\frac{n_{f}}{2}$&$-n_{s}=-n_{f}$\\
\hline
\end{tabular}
\end{center}
Using the above values in the expression (\ref{non}), the
$\sum\limits_{j_{a}}\alpha(j_{a})d(j_{a})$ for the fields in the
adjoint representation as well as
$\sum\limits_{j_{f}}\alpha(j_{f})d(j_{f})$ for the fields in the
fundamental representation vanish. This was also expressed in
\cite{khoze,alvarez}. The remaining part is
\begin{eqnarray}\label{non-2}
\Pi(p)&=&\frac{1}{8\pi^{2}}\int\limits_{0}^{1}dx\
\sum\limits_{j_{a}}\alpha(j_{a})\left(-4C(j_{a})\right)
\left(2K_{0}\left(|\tilde{p}|\sqrt{\Delta}\right)\right)\nonumber\\
&&+\frac{1}{8\pi^{2}}\int \limits_{0}^{1}dx\
\Bigg[\sum\limits_{j_{a}}\alpha(j_{a})(-4C(j_{a}))+\frac{1}{2}\sum\limits_{j_{f}}\alpha(j_{f})(-4C(j_{f}))\Bigg]
\ln\frac{\Delta}{4\pi\mu^{2}},
\end{eqnarray}
leading to
\begin{eqnarray}\label{NS4-yx}
\beta(g;p,\Theta)\equiv -
\frac{g^{3}}{8\pi^{2}}\Bigg[-\frac{n_{f}}{2}+3\int\limits_{0}^{1}dx\
\left(1-2K_{1}(|\tilde{p}|\sqrt{\Delta})
\frac{\tilde{p}^{2}\Delta}{|\tilde{p}|\sqrt{\Delta}}\right)\Bigg].
\end{eqnarray}
Now choosing $n_{f}=2N_{f}$ and performing the same analysis as
above, we arrive at the coefficient of the one-loop
$\beta$-function for $|p|\gg\frac{1}{\sqrt{\Theta}}$ and
$|p|\ll\frac{1}{\sqrt{\Theta}},$\footnote{Note that the
coefficient of the $\beta$-function of noncommutative
supersymmetric ${\cal{N}}=1$ U(1) gauge theory differs from the
coefficient of the $\beta$-function of the commutative
supersymmetric SQCD \cite{peskin} by a factor of 2. This is the
same discrepancy as in the nonsupersymmetric case
\cite{khoze,nakajima}.}
\begin{eqnarray}\label{beta2}
\beta(g;p,\Theta)=\left\{
\begin{array}{cccc}
-\frac{g^{3}}{16\pi^{2}}b_{\ell,N_{f}}\qquad \mbox{with}&b_{\ell,N_{f}}=+2\left(3-N_{f}\right),&&|p|\gg\frac{1}{\sqrt{\Theta}},\\
-\frac{g^{3}}{16\pi^{2}}b_{s,N_{f}}\qquad \mbox{with}&b_{s,N_{f}}=-2\left(3+N_{f}\right),&&|p|\ll\frac{1}{\sqrt{\Theta}}.\\
\end{array}
\right.
\end{eqnarray}
The first expression on the r.h.s. of (\ref{non-2}), is the
contribution from nonplanar diagrams, appearing only in front of
the fields in the adjoint representation. Further, since the
matter fields are in the fundamental representation and therefore
do not receive any nonplanar contributions, the sign of the part
proportional to $N_{f}$ does not change in two limits of
$|p|\gg\frac{1}{\sqrt{\Theta}}$ and
$|p|\ll\frac{1}{\sqrt{\Theta}}$. Our results on the one-loop
$\beta$-function can also be compared with \cite{travaglini} and
the references therein.

According to this result, in $|p|\gg\frac{1}{\sqrt{\Theta}}$
limit, the theory is asymptotically free only for $N_{f}< 3$,
whereas the theory is IR-free in $|p|\ll\frac{1}{\sqrt{\Theta}}$
limit for any arbitrary value of $N_{f}$.
\subsubsection*{Noncommutative $U_{R}(1)$ Anomaly}
Let us concentrate on the anomalies corresponding to the axial
$U_{A}(1)$ and R-symmetry $U_{R}(1)$ transformations
\begin{eqnarray}\label{trafo}
U_{A}(1):\hspace{1cm} Q(x,\theta)\to e^{i\alpha}Q(x,\theta),&\mbox{and}&W_{\alpha}(x,\theta)\to
 W_{\alpha}(x,\theta),\nonumber\\
U_{R}(1):\hspace{0.5cm} Q(x,\theta)\to e^{i\alpha}\
Q(x,e^{-3i\alpha/2}\theta) &\mbox{and}&W_{\alpha}(x,\theta)\to
e^{3i\alpha/2}\ W_{\alpha}(x,e^{-3i\alpha/2}\theta).
\end{eqnarray}
The $U_{R}(1)$ anomaly receives contribution from the $N_{f}$
chiral fermion pair $\psi^{i}_{L}$ and $\tilde{\psi}^{i}_{L}$,
with $i=1,\cdots, N_{f}$ in the matter supermultiplet $Q$ and
$\tilde{Q}$, and from the chiral gaugino $\lambda_{L}$ in the
gauge supermultiplet.\footnote{For convenience, we have taken only
the left handed matter fields $\psi_{L}\equiv
\frac{1+\gamma_{5}}{2}\psi$, $\tilde{\psi}_{L}\equiv
\frac{1+\gamma_{5}}{2}\tilde{\psi}$ and gaugino field
$\lambda_{L}\equiv \frac{1+\gamma_{5}}{2}\lambda$.} Whereas the
chiral fermions $\psi_{L}$ and $\tilde{\psi}_{L}$ transform in the
fundamental and antifundamental representations, respectively, the
gaugino $\lambda_{L}$ transforms in the adjoint representation. In
section 2, we calculated only the $U(1)$ anomaly of matter fields
in the fundamental representation. In this section we will compute
the anomaly arising from gauginos in the adjoint representation.
Surprisingly, this result is also affected by the UV/IR mixing. So
that for vanishing  $|\Theta p|$ the R-anomaly corresponding to
the adjoint gauginos vanishes. In the opposite case, however, {\it
i.e.} if we consider the case of arbitrary but finite $|\Theta
p|$, it appears again and therefore decouple from the R-anomaly
arising from the contribution from fermions in the fundamental
representation. This decoupling is in contrast to ordinary
commutative $N_{c}$ color SQCD.
\par
From section 2, it is simple to find the contribution to the
$U_{R}(1)$ anomaly of the fermions in the {\it fundamental} and
{\it antifundamental} representation. Here we shall focus only on
the nonplanar (invariant) anomaly, because  after the Noether
procedure the only current arising from the {\it global}
$U_{R}(1)$ transformation is the invariant current, whose
corresponding anomaly vanishes for arbitrary but finite $|\Theta
p|$. When we take the limit $|\Theta p|\to 0$, its value is given
by
\begin{eqnarray}
{\cal{A}}_{\psi}=2N_{f}\alpha R(\psi) \left(-\frac{1}{32\pi^{2}}
F_{\mu\nu}\star' \tilde{F}^{\mu\nu}+\cdots\right),
\end{eqnarray}
where $R(\psi)$ is the $R$-charge of the chiral fermion
$\psi_{L}$. The extra terms are the contributions of the open
Wilson line.
\par
To find the contribution to $U_{R}(1)$ anomaly corresponding to
the chiral gaugino field $\lambda_{L}$, which in contrast to the
matter fields are in the {\it adjoint} representation, we follow
the Fujikawa method outlined in section 2. The partition function
corresponding to the gauginos in the adjoint representation is
given by
\begin{eqnarray}
{\cal{Z}}=\int {\cal{D}}\lambda_{L}\ {\cal{D}}\bar{\lambda}_{L}\
e^{-iS_{\lambda}[\lambda_{L},\bar{\lambda}_{L}]},
\end{eqnarray}
with $S_{\lambda}=\int d^{4}x\ {\cal{L}}_{\lambda}$, and the
Lagrangian density
\begin{eqnarray}\label{ladj}
{\cal{L}}_{\lambda}\equiv i\bar{\lambda}_{L}(x)\ \DS\
\lambda_{L}(x),
\end{eqnarray}
where the covariant derivative acts on the gaugino field in the
adjoint representation
\begin{eqnarray}
D_{\mu}\lambda_{L}(x)=\partial_{\mu}\lambda_{L}(x)+ig\big[A_{\mu}(x),\lambda_{L}(x)\big]_{\star}.
\end{eqnarray}
Under the antifundamental local change of variables
\begin{eqnarray}
\delta
\lambda_{L}(x)=iR(\lambda)\lambda_{L}(x)\star\alpha(x),\hspace{1cm}\mbox{and}\hspace{1cm}
\delta\bar{\lambda}_{L}(x)=-iR(\lambda)\alpha(x)\star\bar{\lambda}_{L}(x),
\end{eqnarray}
the action transforms as
\begin{eqnarray}
\delta S_{\lambda}= -\int d^{4}x\
D_{\mu}J^{\mu}_{\lambda}(x)\star\alpha(x),\qquad\mbox{with}\qquad
J_{\lambda}^{\mu}=\bar{\lambda}_{L}\gamma^{\mu}\lambda_{L}.
\end{eqnarray}
Further using the Jacobian  of the transformation we have
\begin{eqnarray}
\lefteqn{ {\cal{D}}\lambda_{L}\
{\cal{D}}\bar{\lambda}_{L}\longrightarrow}\nonumber\\
&& \longrightarrow{\cal{D}}\lambda'_{L}\
{\cal{D}}\bar{\lambda'}_{L}= \exp\left(-2i R(\lambda)\int d^{4}x\
\alpha(x)\star\sum\limits_{n}\varphi^{\dagger}_{n,\alpha}(x)(\gamma_{5})^{\alpha\beta}\star\varphi_{n,\beta}(x)
\right){\cal{D}}\lambda_{L}\ {\cal{D}}\bar{\lambda}_{L}.\nonumber\\
\end{eqnarray}
The invariance of the partition function leads therefore to the
anomaly corresponding to the gaugino field
\begin{eqnarray}
{\cal{A}}_{\lambda}=\lim\limits_{M\to \infty}\ 2\
R(\lambda)\sum\limits_{n}\varphi^{\dagger}_{n,\alpha}(x)\left(\gamma_{5}\right)^{\alpha\beta}\
\left(e^{-\frac{\DS^{2}}{M^{2}}}\right)_{\beta}^{\delta}\
\star\varphi_{n,\delta}(x) ,
\end{eqnarray}
where we have introduced the regulator $M$ in the damping factor
$\exp\left(-\DS^{2}/M^{2}\right)$. Using the definition
$\DS^{2}=D_{\mu}D^{\mu}+\frac{i}{2}\sigma^{\mu\nu}[D_{\mu},
D_{\nu}]$, noting that the functions $\varphi_{n}$'s are in the
adjoint representation and using the completeness of
$\varphi_{n}$'s in the Fourier space, we arrive at
\begin{eqnarray}
{\cal{A}}_{\lambda}&=&\lim\limits_{M\to \infty}\ 2\ R(\lambda)\
 \int\frac{d^{4}k}{(2\pi)^{4}}\ e^{\frac{k^{2}}{M^{2}}}e^{-ikx}\star\mbox{tr}\left(\gamma_{5}\
  \exp\left(\frac{i}{2M^{2}}\sigma^{\mu\nu}[D_{\mu}, D_{\nu}]\right)\right)\star e^{ikx}\nonumber\\
&=&\lim\limits_{M\to \infty}\frac{ig^{2}}{M^{4}}\ R(\lambda)\
\varepsilon^{\mu\nu\rho\lambda}\ \int\frac{d^{4}k}{(2\pi)^{4}}\
e^{\frac{k^{2}}{M^{2}}} e^{-ikx}\
\star\big[F_{\mu\nu}(x),\big[F_{\rho\lambda}(x),e^{ikx}\big]_{\star}\big]_{\star}.
\end{eqnarray}
Using further the relation (\ref{n1}), we arrive at
\begin{eqnarray}\label{k1}
\lefteqn{ {\cal{A}}_{\lambda}=\lim\limits_{M\to
\infty}\frac{ig^{2}}{M^{4}}\ R(\lambda)\
\varepsilon^{\mu\nu\rho\lambda}\ \int\frac{d^{4}k}{(2\pi)^{4}}\
e^{+\frac{k^{2}}{M^{2}}}
}\nonumber\\
&&\times \left(F_{\mu\nu}(x-\tilde{k})\star
F_{\rho\lambda}(x-\tilde{k})-2 F_{\mu\nu}(x-\tilde{k})\star
F_{\rho\lambda}(x)+F_{\mu\nu}(x)\star F_{\rho\lambda}(x)\right).
\end{eqnarray}
The three terms appearing on the r.h.s. of (\ref{k1}) are
evaluated separately. After integrating over $k$, the first term,
for instance, leads to
\begin{eqnarray}
\lefteqn{\hspace{-.7cm} \lim\limits_{M\to
\infty}+\frac{ig^{2}}{M^{4}}\ R(\lambda) \int
\frac{d^{4}k}{(2\pi)^{4}}\frac{d^{4}k_{1}}{(2\pi)^{4}}\frac{d^{4}k_{2}}{(2\pi)^{4}}\
e^{+\frac{k^{2}}{M^{2}}-2i(k_{1}+k_{2})\times k}\
F_{\mu\nu}(k_{1})\ e^{ik_{1}x}\star e^{ik_{2}x}\
\tilde{F}^{\mu\nu}(k_{2})
}\nonumber\\
&=&\lim\limits_{M\to \infty}-\frac{g^{2}}{16\pi^{2}}\ R(\lambda)
\int \frac{d^{4}k_{1}}{(2\pi)^{4}}\frac{d^{4}k_{2}}{(2\pi)^{4}}\
e^{-\frac{M^{2}}{4}(k_{1}+k_{2})\circ (k_{1}+k_{2})}
F_{\mu\nu}(k_{1})\ e^{ik_{1}x}\star e^{ik_{2}x}\ \tilde{F}^{\mu\nu}(k_{2}).\nonumber\\
\end{eqnarray}
As for the invariant anomaly, a UV/IR mixing occurs here.
Following the same arguments as in the previous section, and
calculating the contributions of other two terms, it turns out the
anomaly corresponding to the gauginos in the adjoint
representation vanishes in the limit of vanishing $|\Theta p|$,
and for arbitrary but finite $|\Theta p|$, it is given by
\begin{eqnarray}
{\cal{A}}_{\lambda}= -\frac{g^{2}}{16\pi^{2}}\ R(\lambda)\
F_{\mu\nu}\star \tilde{F}^{\mu\nu}.
\end{eqnarray}
Let us now summarize our results about the anomalies corresponding
to $U_{A}(1)$ and $U_{R}(1)$ symmetries. As we have seen in the
limit of vanishing  $|\Theta p|$, $\delta_{A}{\cal{L}}$ is given
by
\begin{eqnarray}\label{ncaxial}
\delta_{A}{\cal{L}}= 2N_{f}\alpha
\left(-\frac{1}{32\pi^{2}}F_{\mu\nu}\star'
\tilde{F}^{\mu\nu}+\cdots\right),
\end{eqnarray}
and $\delta_{R}{\cal{L}}$ by
\begin{eqnarray}\label{ncrsym}
\delta_{R}{\cal{L}}= 2N_{f}\alpha
R(\psi)\left(-\frac{1}{32\pi^{2}}F_{\mu\nu}\star'
\tilde{F}^{\mu\nu}+\cdots\right).
\end{eqnarray}
In (\ref{ncaxial}) and (\ref{ncrsym}) the ellipses denote the
higher order terms in the expansion of the open Wilson line. For
an arbitrary but finite $|\Theta p|$, however,
$\delta_{A}{\cal{L}}$ vanishes and $\delta_{R}{\cal{L}}$ is given
by
\begin{eqnarray}\label{ncrsymhigh}
\delta_{R}{\cal{L}}= 2\alpha
R(\lambda)\left(-\frac{1}{32\pi^{2}}F_{\mu\nu}\star
\tilde{F}^{\mu\nu}\right).
\end{eqnarray}
The above results will be used in the next section to calculate
the effective superpotential of the noncommutative supersymmetric
theory. Before proceeding let us describe our strategy:
\par
As was mentioned before, the full effective superpotential of the
theory consists of two parts, the dynamical
Veneziano-Yanckielowicz (VY) superpotential \cite{vy,vyt}, and the
tree level superpotential. We will determine the noncommutative VY
superpotential, as in the ordinary commutative case \cite{vyt},
using the differential equations assuring that the symmetries of
the original theory are also preserved in the effective theory. To
do this, we use the anomalies corresponding to $U_{A}(1)$ and
$U_{R}(1)$ symmetries and will first determine the relevant
degrees of freedom for the effective theory, and then, using the
corresponding selection rules, we arrive at the VY effective
superpotential.
\par
As for the anomalies, we have seen that, due to UV/IR mixing, a
singularity appears in the limit of vanishing $|\Theta p|$. To
begin, in section 4.2, we will first consider the case of
vanishing $|\Theta p|$. In this case the relevant degree of
freedom of the effective theory, apart from the meson field
$T\equiv \tilde{Q}\star Q$, is the gauge {\it invariant} Konishi
anomaly $S'\propto W_{\alpha}\star' W^{\alpha}+\cdots$, which
contains the {\it invariant} anomaly
$F_{\mu\nu}\star'\tilde{F}^{\mu\nu}+\cdots$ as a component. Here
the effective superpotential will be determined using the
selection rules from the corresponding {\it invariant} anomalies,
and from the coefficient of the one-loop $\beta$-function in
$|p|\ll\frac{1}{\sqrt{\Theta}}$ limit {\it i.e.} $b_{s,N_{f}}$
from (\ref{beta2}). This program ends up with an effective
superpotential for the gauge {\it invariant} superfield $S'$ and
the meson field $T$. We will integrate out each of these
superfields to find first the effective superpotential for $S'$
and then the Affleck-Dine-Seiberg (ADS) effective superpotential
\cite{seiberg1,seiberg,seiberg2} for $T$.
\par
We then continue to determine the effective superpotential for
arbitrary but finite $|\Theta p|$ in section 4.3. According to the
results for the anomalies corresponding to $U_{A}(1)$ and
$U_{R}(1)$ symmetries, the relevant degrees of freedom are the
meson superfield $T$, and the gauge {\it covariant} Konishi
anomaly $S\propto W_{\alpha}\star W^{\alpha}$, which contains the
gauge {\it covariant} anomaly $F_{\mu\nu}\star\tilde{F}^{\mu\nu}$
as a component. Using these anomalies and the coefficient of the
one-loop  $\beta$-function, $b_{\ell,N_{f}}$ from (\ref{beta2}),
in $|p|\gg\frac{1}{\sqrt{\Theta}}$ limit, we will determine the
effective superpotential for the gauge {\it covariant} superfield
$S$ and the meson field $T$. Again the effective superpotentials
will be determined for $S$ and $T$ separately.
\subsubsection*{4.2\hspace{0.3cm}Case 1: Effective superpotential for $S'$}
Here we consider the limit of vanishing $|\Theta p|$. According to
our results from the anomalies corresponding to $U_{A}(1)$ and
$U_{R}(1)$ symmetries in this limit, (\ref{ncaxial}) and
(\ref{ncrsym}), the relevant degrees of freedom in this case are
the invariant (nonplanar) Konishi anomaly
\begin{eqnarray*}
S'\equiv -S_{inv.}^{nonplanar}=+\frac{1}{32\pi^{2}}\
W_{\alpha}\star' W^{\alpha}+\cdots,
\end{eqnarray*}
with the extra terms denoting the corrections arising from the
Wilson line attachment, and the meson field $T_{ij}\equiv
\tilde{Q}_{i}\star Q_{j}$. The effective action is therefore given
by
\begin{eqnarray}\label{eff.}
I_{eff.}=\int d^{2}\theta\ d^{4}x\ W_{dyn.}(T, S').
\end{eqnarray}
As indicated in section 2, we will find $W_{dyn.}$ in term of $S'$
without necessarily committing ourselves to the particular form
above. Varying this effective action with respect to the
$U_{A}(1)$ and $U_{R}(1)$ transformations and using the results
from (\ref{ncaxial}) and (\ref{ncrsym}) with $R(\psi)=-1/2$, we
arrive at two differential equations determining $W_{dyn.}$
uniquely
\begin{eqnarray}\label{a1}
T\ \frac{\partial W_{dyn.}}{\partial T}=-N_{f}S',
\end{eqnarray}
and
\begin{eqnarray}\label{a3}
-W_{dyn.}+S'\ \frac{\partial W_{dyn.}}{\partial S'}+\frac{2}{3}\
T\frac{\partial W_{dyn.}}{\partial T}=+\frac{N_{f}}{3} S'.
\end{eqnarray}
Putting the first equation (\ref{a1}) in (\ref{a3}), it is given
by
\begin{eqnarray}\label{a2}
-W_{dyn.}+S'\ \frac{\partial W_{dyn.}}{\partial S'}-N_{f} S'=0.
\end{eqnarray}
The solution to the differential equations (\ref{a1}) and
(\ref{a2}) reads therefore
\begin{eqnarray}\label{dyn1}
W_{dyn.}(T, S'; m,\lambda;
\Lambda_{N_{f}},\Lambda_{\Theta})=S'\left(\log\left(\frac{S'^{+N_{f}}}{\Lambda_{\Theta}^{\alpha}\
\Lambda_{N_{f}}^{\kappa}\det T}\right)-N_{f}\right).
\end{eqnarray}
To determine the exponent $\alpha$ and $\kappa$ of the holomorphic
scale $\Lambda_{N_{f}}$ and the new mass scale
$\Lambda_{\Theta}\equiv \frac{1}{\sqrt{\Theta}}$, we use the
selection rules as in the ordinary commutative SQCD
\cite{seiberg2, terning}. Requiring that the effective
superpotential is invariant under $U_{R}(1)$ and $U_{A}(1)$
transformations, and using the results from (\ref{ncaxial}) and
(\ref{ncrsym}), as well as the coefficient of the one-loop
$\beta$-function in the $|p|\ll\Lambda_{\Theta}$ limit, {\it i.e.}
$b_{s,N_{f}}$ from (\ref{beta2}), the axial and $R$-charges and
the mass dimension of the quantities appearing in the dynamical
superpotential are determined. Table 1 summarizes these results.
\begin{table}[hbt]
\begin{center}
\begin{tabular}{|c||c|c|c|}
\hline
&$U_R(1)$-charge&$U_A(1)$-charge&$m$-dim\\
\hline
det$T$&$3N_{f}$&$2N_{f}$&$2N_{f}$\\
$\Lambda_{\Theta}$&$0$&$0$&$+1$\\
$\left(\Lambda_{N_f}\right)^{b_{{s},N_{f}}}$&$0$&$2N_{f}$&$b_{s,N_{f}}$\\
$S'$&$3$&0&3\\
\hline
$W_{dyn.}$&$3$&0&3\\
\hline
\end{tabular}
\caption{$U_{R}(1)$, $U_{A}(1)$ and mass dimensions for case 1.}
\end{center}
\end{table}

The values $\kappa=-b_{s,N_{f}}$ and $\alpha=N_{f}+b_{s,N_{f}}$
with $b_{s,N_{f}}=-2(3+N_{f})$ guarantee that the argument of
logarithm in (\ref{dyn1}) is dimensionless and has vanishing R-
and axial charges. The dynamical superpotential for the
superfields $S'$ and $T$ reads therefore
\begin{eqnarray}\label{ncsol1}
W_{dyn.}( T,S';\Lambda_{N_{f}}, \Lambda_{\Theta})=+S'\left(\log\
\left(\frac{S'^{N_{f}}\Lambda_{\Theta}^{+(N_{f}+6)}}
{\Lambda_{N_{f}}^{+2(3+N_{f})}\det T}\right)-N_{f}\right).
\end{eqnarray}
Adding, as in the ordinary case, the tree level superpotential
\begin{eqnarray}\label{ncwtree}
W_{tree}=m\ \mbox{tr}\ T+\lambda\ \mbox{tr}\ T\star T,
\end{eqnarray}
with $\star$-products replacing the ordinary products, to the
dynamical superpotential (\ref{ncsol1}), the full effective
superpotential reads
\begin{eqnarray}\label{ncsol2}
\lefteqn{W_{eff.}(T, S';m,\lambda;
\Lambda_{N_{f}},\Lambda_{\Theta})=
}\nonumber\\
&=& m\ \mbox{tr}\ T+\lambda\ \mbox{tr}\ T^{2}+S'\left(\log\
\left(\frac{S'^{N_{f}}\Lambda_{\Theta}^{+(N_{f}+6)}}{\Lambda_{N_{f}}^{+2(3+N_{f})}\det
T}\right)- N_{f}\right).
\end{eqnarray}
Integrating out $S'$ the Affleck-Dine-Seiberg (ADS) superpotential
of the theory is given by
\begin{eqnarray}
W_{eff.}^{ADS}=m\ \mbox{tr}\ T+\lambda\ \mbox{tr}\
T^{2}-N_{f}\left(\frac{\Lambda_{N_{f}}^{-2(3+N_{f})}\
\Lambda_{\Theta}^{+(N_{f}+6)}}{\det T}\right)^{-\frac{1}{N_{f}}}.
\end{eqnarray}
Further as in the ordinary case, integrating the meson field $T$
from (\ref{ncsol2}) leads to the VY superpotential for the pure
gauge theory as a function of $S'$. Doing this, we arrive first at
the noncommutative Konishi equation
\begin{eqnarray}
m\ <T_{i}>+2\lambda\ <T_{i}^{2}>=S',
\end{eqnarray}
where  $T_{i}'s, i=1,\cdots N_{f}$ are the diagonal elements of
the meson matrix.  This equation can be solved using the
factorization $<T_{i}^{2}>=<T_{i}>^{2}$ as in the ordinary
commutative SQCD theory \cite{oz}, to yield the noncommutative
meson field $<\mbox{tr}\ T>$ as a function of $S'$. Choosing
$N_{f}^{+}$ eigenvalues to reduce to $T=0$ and $N_{f}^{-}$
eigenvalues to $T=-m/2\lambda$, the classical vacua of the theory,
we arrive at \cite{oz, wheater}
\begin{nedalph}\label{eigen}
<\mbox{tr}\ T>=N_{f}^{+}\left(-\frac{m}{4\lambda}+
\frac{m}{4\lambda}\sqrt{1+\frac{8\lambda
S'}{m^2}}\right)+N_{f}^{-}\left(-\frac{m}{4\lambda}-
\frac{m}{4\lambda}\sqrt{1+\frac{8\lambda S'}{m^2}}\right),
\end{eqnarray}
and
\begin{eqnarray}
<\mbox{tr}\ T^{2}>=N_{f}^{+}\left(-\frac{m}{4\lambda}+
\frac{m}{4\lambda}\sqrt{1+\frac{8\lambda
S'}{m^2}}\right)^2+N_{f}^{-}\left(-\frac{m}{4\lambda}-
\frac{m}{4\lambda}\sqrt{1+\frac{8\lambda S'}{m^2}}\right)^2.
\end{nedalph}

Now we have to match the RG-invariant scale $\Lambda_{N_{f}}$ and
$\Lambda_{\Theta}$ defined for a theory with $N_{f}$ massless
flavors, to a new combination for a theory without massless
flavors, {\it i. e.} a pure gauge theory. To do this we begin by
adding the mass term $m\ \mbox{tr}\ T$ to the dynamical ADS
superpotential of a theory with $N_{f}$ massless flavors
\begin{eqnarray}
W_{eff}=m\ \mbox{tr}\
T-N_{f}\left(\frac{\Lambda_{N_{f}}^{-2(3+N_{f})}\
\Lambda_{\Theta}^{+(N_{f}+6)}}{\det T}\right)^{-\frac{1}{N_{f}}}.
\end{eqnarray}
Integrating out only $N_{f}-n_{f}$ massive flavors from $N_{f}$
flavors, we arrive at the superpotential for a theory with $n_{f}$
massless flavors
\begin{eqnarray}
W_{eff}=-n_{f}\left(\frac{m^{(N_{f}-n_{f})}
\Lambda_{N_{f}}^{-2(3+N_{f})}\ \Lambda_{\Theta}^{+(N_{f}+6)}}{\det
\hat{T}}\right)^{-\frac{1}{n_{f}}}.
\end{eqnarray}
Here $\hat{T}$ is the $n_{f}\times n_{f}$ meson matrix built from
$n_{f}$ massless flavors.  Comparing now this superpotential with
the dynamical ADS superpotential with $n_{f}$ massless flavors
\begin{eqnarray}
W_{eff}=-n_{f}\left(\frac{\hat{\Lambda}_{n_{f}}^{-2(n_{f}+3)}\hat{\Lambda}_{\Theta}^{+(n_{f}+6)}}{\det
\hat{T}}\right)^{-\frac{1}{n_{f}}},
\end{eqnarray}
we arrive at the following consistent scale matching
\begin{eqnarray}
\hat{\Lambda}_{n_{f}}^{-2(n_{f}+3)}\hat{\Lambda}_{\Theta}^{+(n_{f}+6)}=
m^{N_{f}-n_{f}}
\Lambda_{N_{f}}^{-2(N_{f}+3)}\Lambda_{\Theta}^{+(N_{f}+6)}.
\end{eqnarray}
This matching equation can be now used to define the scale for a
pure gauge theory, indicated by $n_{f}=0$
\begin{eqnarray}\label{match}
\frac{\hat{\Lambda}_{\Theta}}{\hat{\Lambda}_{0}}= \left(m^{N_{f}}
\Lambda_{N_{f}}^{-2N_{f}}\Lambda_{\Theta}^{N_{f}}\right)^{\frac{1}{6}}\
\frac{\Lambda_{\Theta}}{\Lambda_{N_{f}}}.
\end{eqnarray}
Inserting (\ref{eigen}-b) and (\ref{match}) in the effective
superpotential (\ref{ncsol2}), we arrive at the effective
superpotential of noncommutative ${\cal{N}}=1$ supersymmetric
$U(1)$ gauge theory with all $N_{f}$ massive flavors integrated
out
\begin{eqnarray}\label{ncsol4}
W_{eff.}(S'; m,\lambda;\hat{\Lambda}_{0},
\hat{\Lambda}_{\Theta})&=& +6
S'\log\frac{\hat{\Lambda}_{\Theta}}{\Lambda_{0}} -\frac{N_{f}}{2}
S' -N_{f}\frac{m^{2}}{8\lambda}
+(N_{f}^{+}-N_{f}^{-})\frac{m^{2}}{8\lambda}\sqrt{1+\frac{8\lambda
S'}{m^{2}}}
\nonumber\\
&&
+S'\log\Bigg[\left(\frac{1}{2}+\frac{1}{2}\sqrt{1+\frac{8\lambda
S'}{m^{2}}}\right)^{N_{f}^{+}}\
 \left(\frac{1}{2}-\frac{1}{2}\sqrt{1+\frac{8\lambda S'}{m^{2}}}\right)^{N_{f}^{-}}\Bigg].\nonumber\\
\end{eqnarray}
This result can be compared with the effective superpotential of
ordinary commutative $N_{c}$ colors SQCD which depends on the
commutative gaugino bilinear $S\propto
\mbox{tr}(W_{\alpha}W^{\alpha})$ \cite{oz}
\begin{eqnarray}\label{sol4}
W_{eff.}&=&-N_{c}
S\left(\log\frac{S}{\hat{\Lambda}_{0}^3}-1\right)-
\frac{N_{f}}{2}S
-N_{f}\frac{m^{2}}{8\lambda}+(N_{f}^{+}-N_{f}^{-})\frac{m^{2}}{8\lambda}\sqrt{1+\frac{8\lambda S}{m^2}}\nonumber\\
&& +S\log\Bigg[
\left(\frac{1}{2}+\frac{1}{2}\sqrt{1+\frac{8\lambda
S}{m^{2}}}\right)^{N_{f}^{+}}\
\left(\frac{1}{2}-\frac{1}{2}\sqrt{1+\frac{8\lambda
S}{m^{2}}}\right)^{N_{f}^{-}}\Bigg].
\end{eqnarray}
Although an apparent similarity between these two superpotentials
exists, the noncommutative superpotential (\ref{ncsol4}) is a
nontrivial function of gauge field supermultiplet appearing in the
Wilson line attachment of $S'$.
\subsubsection*{4.3\hspace{0.3cm}Case 2: Effective superpotential for $S$}
Here we consider the case of arbitrary but finite $|\Theta p|$.
The relevant degrees of freedom in this case are the covariant
(planar) Konishi anomaly
\begin{eqnarray*}
S\equiv -S_{cov.}^{planar}=+\frac{1}{32\pi^{2}}\ W_{\alpha}\star
W^{\alpha},
\end{eqnarray*}
 and the meson field $T_{ij}\equiv \tilde{Q}_{i}\star
Q_{j}$. The effective action can therefore be given by
\begin{eqnarray}\label{eff2}
I_{eff.}=\int d^{2}\theta\ d^{4}x\ W_{dyn.}(T, S).
\end{eqnarray}
Using the corresponding results $\delta_{A}{\cal{L}}=0$
 and $\delta_{R}{\cal{L}}$ from (\ref{ncrsymhigh}) with
$R(\lambda)=3/2$, we arrive at two differential equations
determining $W_{dyn.}$ uniquely,
\begin{eqnarray}\label{a11}
T\ \frac{\partial W_{dyn.}}{\partial T}=0,
\end{eqnarray}
and
\begin{eqnarray}\label{a12}
-W_{dyn.}+S\ \frac{\partial W_{dyn.}}{\partial S}+\frac{2}{3}\ T\
\frac{\partial W_{dyn.}}{\partial T}=-S.
\end{eqnarray}
Plugging the first equation (\ref{a11}) in (\ref{a12}), the second
equation reads
\begin{eqnarray}\label{a21}
-W_{dyn.}+S\ \frac{\partial W_{dyn.}}{\partial S}+S=0,
\end{eqnarray}
which can be solved to yield
\begin{eqnarray}
W_{dyn.}\left(T,S;\Lambda_{N_{f}},\Lambda_{\Theta}\right)=
S\left(\log\left(\frac{S^{-1}}{\Lambda_{\Theta}^{\alpha}\Lambda_{N_{f}}^{\kappa}}\right)+1\right).
\end{eqnarray}
To determine the two exponent $\alpha$ and $\kappa$, we require
the invariance of the effective superpotential with respect to
$U_{A}(1)$ and $U_{R}(1)$ transformations, and use the coefficient
of the one-loop $\beta$-function in the $|p|\gg\Lambda_{\Theta}$
limit, {\it i.e.} $b_{\ell,N_{f}}$ from (\ref{beta2}). This is in
fact admissible since in this case, $|\Theta p|$ is arbitrary but
finite. The R-charge and the axial charge of the quantities
appearing in the superpotential are therefore determined. Table 2
summarizes these results.

\begin{table}[tbh]
\begin{center}
\begin{tabular}{|c||c|c|c|}
\hline
&$U_R(1)$-charge&$U_A(1)$-charge&$m$-dim\\
\hline
det$T$&$3N_{f}$&$2N_{f}$&$2N_{f}$\\
$\Lambda_{\Theta}$&$0$&$0$&$+1$\\
$\left(\Lambda_{N_f}\right)^{b_{\ell, N_{f}}}$&$3$&$0$&$b_{\ell, N_{f}}$\\
$S$&$3$&0&3\\
\hline
$W_{dyn.}$&$3$&0&3\\
\hline
\end{tabular}
\end{center}
\caption{$U_{R}(1)$, $U_{A}(1)$ and mass dimension for case 2.}
\end{table}

The exponent $\alpha$ and $\kappa$ in the dynamical superpotential
can be determined correspondingly. They are given by
$\kappa=-b_{\ell,N_{f}}=-2(3-N_{f})$ from the coefficient of the
one-loop $\beta$-function in the $|p|\gg \Lambda_{\Theta}$ limit,
and $\alpha=b_{\ell,N_{f}}-3=3-2N_{f}$. The dynamical
superpotential as a function of $S$ reads therefore
\begin{eqnarray}\label{ncsol11}
W_{dyn.}(S;\Lambda_{N_{f}},\Lambda_{\Theta})=S\left(\log\
\left(\frac{S^{-1}\
\Lambda_{\Theta}^{2N_{f}-3}}{\Lambda_{N_{f}}^{-2(3-N_{f})}}\right)+1\right).
\end{eqnarray}
The full effective superpotential in this case is given by taking
(\ref{ncwtree}) and adding it to the noncommutative dynamical
superpotential (\ref{ncsol11})
\begin{eqnarray}\label{ncsol21}
W_{eff.}(T, S; m,\lambda; \Lambda_{N_{f}},\Lambda_{\Theta})=m\
\mbox{tr}\ T+\lambda\ \mbox{tr}\ T^{2}-S\left(\log\ \left(\frac{S\
\Lambda_{\Theta}^{3-2N_{f}}}{\Lambda_{N_{f}}^{+2(3-N_{f})}}\right)-1\right).
\end{eqnarray}
Integrating out the $S$ field, we arrive at the ADS effective
superpotential
\begin{eqnarray}
W_{eff.}^{ADS}=m\ \mbox{tr}\ T+\lambda\ \mbox{tr}\
T^{2}+\Lambda_{N_{f}}^{6-2N_{f}}\ \Lambda_{\Theta}^{2N_{f}-3}.
\end{eqnarray}
Further integrating out the meson field $T$ from the full
superpotential (\ref{ncsol21}), we arrive first at
\begin{eqnarray}
m+2\lambda\ (T)_{i}=0,
\end{eqnarray}
which can be solved to yield the noncommutative meson field
$<\mbox{tr}\ T>=-\frac{N_{f}m}{2\lambda}$. Replacing these results
back in the full effective superpotential (\ref{ncsol21}) and
noting that $<\mbox{tr}\ T^{2}>=\frac{N_{f}m^{2}}{4\lambda^{2}}$,
we have
\begin{eqnarray}
W_{eff.}(<\mbox{tr}\ T>=-\frac{N_{f}m}{2\lambda}; m,\lambda;
\hat{\Lambda}_{0},\hat{\Lambda}_{\Theta})
=-\frac{N_{f}m^{2}}{4\lambda}- S\left(\log
\frac{S\hat{\Lambda}_{\Theta}^{3}}{\hat{\Lambda}_{0}^{6}}-1\right),
\end{eqnarray}
where the scale matching
$
\frac{ \hat{ \Lambda}_{0}^{6}}{\hat{\Lambda}_{\Theta}^{3}}=
\frac{\Lambda_{N_{f}}^{6-2N_{f}}}{\Lambda_{\Theta}^{3-2N_{f}} },
$
is used.
\subsubsection*{5\hspace{0.3cm}Discussion}\label{Sec5}
In the first part of this work, because of the significant role
that subtleties of the nonplanar contributions to the anomaly play
in the effective action, planar and nonplanar anomalies of
noncommutative $U(1)$ gauge theory with matter fields in the
fundamental representation are calculated anew using Fujikawa's
path integral method \cite{ned1, ned2}.
\par
In the second part of this work, the Konishi anomalies of
noncommutative  ${\cal{N}}=1$ super\-symmetric $U(1)$ gauge
theory, with bosonic noncommutativity are calculated.  As in the
nonsupersymmetric case, the invariant (nonplanar) Konishi anomaly
involves a supersymmetric noncommutative open Wilson line with a
length proportional to the noncommutative length scale
$\sqrt{\Theta}$ and exhibits therefore the same UV/IR mixing.
\par
Finally, in the last part of this work, the exact effective
superpotential for noncommutative ${\cal{N}}=1$ supersymmetric
$U(1)$ gauge theory with $N_{f}$ flavors in the fundamental and
$N_{f}$ in antifundamental representation  is calculated in terms
of the relevant degrees of freedom, the meson field $T_{ij}\equiv
\tilde{Q}_{i}\star Q_{j}$ and the superfields $S\propto
W_{\alpha}\star W^{\alpha}$ and $S'\propto W_{\alpha}\star'
W^{\alpha}+\cdots$, where the extra terms are the contributions
from the expansion of Wilson line.
\par
The VY superpotential for the superfields $S$ and $S'$ as well as
the ADS superpotential for the meson fields $T$ are affected by
UV/IR mixing which has its origins both in their dependence on the
planar and nonplanar anomalies and in their $\beta$-function
dependence. We note that, in contrast to the ordinary SQCD, the
contribution of the gauginos in the adjoint representation to the
$U_{R}(1)$ anomaly appears only for arbitrary but finite
noncommutative momenta, whereas the chiral fermions in the
fundamental representation contribute only in the limit of
vanishing $|\Theta p|$ to the $U_{R}(1)$ anomaly.

A novel feature of the present work is the appearance of new scale
$\Lambda_{\Theta}\equiv \frac{1}{\sqrt{\Theta}}$ in the
superpotential. In particular the noncomomutative version of
Affleck-Dine-Seiberg superpotential of the meson fields depends on
this noncommutative mass scale. The noncommutative scale appears
also elsewhere in the literature \cite{khoze, chu}. We must
emphasize that although the dependence of the effective action on
the superfield $S'$ and $S$ is familiar and, up to some numerical
factors, similar to the ordinary $N_{c}$ colors SQCD \cite{oz},
the dependence of $S'$ on the gauge field supermultiplet arising
from the Wilson line is highly nontrivial.
\subsubsection*{6\hspace{0.3cm}Acknowledgment}
We would like to thank M. Alishahiha for collaboration at the
early stages of this work. We are also grateful to H. Arfaei and
S. Parvizi for discussions. F. A. is indebted to R. Banerjee for
discussions.

\end{document}